\documentclass[11pt,a4paper]{article}
\usepackage{jheparxiv}
\usepackage{amsmath}
\usepackage{amsfonts}
\usepackage{amssymb}
\usepackage{latexsym}
\usepackage{mathrsfs}
\usepackage{braket}		
\usepackage{color}
\usepackage{xcolor}
\usepackage{slashed}
\usepackage{twistor}
\usepackage[all]{xy}
\usepackage{tikz-cd}
\usepackage{mathtools}
\usepackage{bm, bbm}
\usepackage{color}
\usepackage{braket}
\usepackage{float}
\usepackage[]{todonotes}
\usepackage{tikz-feynhand}
\usetikzlibrary{decorations.pathreplacing}
\usepackage{subcaption}
\usepackage{tcolorbox}

\makeatletter
\renewcommand{\todo}[2][]{%
    \@todo[caption={#2}, #1]{\begin{spacing}{0.5}#2\end{spacing}}%
} 
\makeatother 
\usepackage{setspace}

\newcommand{\dd}{{\rm d}}
\newcommand{\nn}{\nonumber \\}

\newcommand{\Amp}{ \mathcal{A}}

\newcommand{\hd}{\hat{\rm d}}
\newcommand{\hdelta}{\hat{\rm \delta}}

\newcommand{\bmp}{\bm{p}}
\newcommand{\bmP}{\bm{P}}
\newcommand{\bmq}{\bm{q}}
\newcommand{\bmk}{\bm{k}}
\newcommand{\bmx}{\bm{x}}
\newcommand{\bmX}{\bm{X}}
\newcommand{\bmpb}{\bm{p}_{\rm bulk}}
\newcommand{\bmVb}{\bm{V}_{\rm bulk}}
\newcommand{\bmr}{\bm{\rho}}
\newcommand{\bml}{\bm{\ell}}

\newcommand{\eqcl}{\overset{{\rm BO}}{=}}
\newcommand{\eqcons}{\underset{{\rm cons}}{=}}
\newcommand{\eqcc}{\overset{\rm BO}{\underset{\rm cons}{=}}}

\newcommand{\tbox}[2]{
\begin{tcolorbox}[ams align,colback=black!5!white,colframe=black!75!white,title=#1]
#2
\end{tcolorbox}
\noindent
}

\tcbuselibrary{theorems}

\newcommand{\twoL}[3]{
\draw (#1, #3/2) -- (#2, #3/2);
\draw (#1, -#3/2) -- (#2, -#3/2);
}
\newcommand{\threeL}[3]{
\draw (#1, 0) -- (#2, 0);
\draw (#1, -#3) -- (#2, -#3);
\propag[boson] (#1, #3) -- (#2, #3);
}
\newcommand{\twoLlow}[3]{
\draw (#1, 0) -- (#2, 0);
\draw (#1, -#3) -- (#2, -#3);
}
\newcommand{\threeLlow}[3]{
\draw (#1, -#3/3) -- (#2, -#3/3);
\draw (#1, -#3) -- (#2, -#3);
\propag[boson] (#1, #3/3) -- (#2, #3/3);
}
\newcommand{\bub}[5]{
\filldraw [fill=white] (#1,#2) circle [radius=#3];
\node at (#1,#2) {#5 $#4$};
}

\newcommand{\Atwo}[1]{
\begin{tikzpicture}[baseline=-2]
\begin{feynhand}
\twoL{-0.8}{0.8}{0.3}
\bub{0}{0}{1.5*0.3}{#1}{}
\end{feynhand}
\end{tikzpicture}
}
 
\newcommand{\Arad}[1]{
\begin{tikzpicture}[baseline=-2]
\begin{feynhand}
\def\y{0.3}
\def\segmentLength{1}
\pgfmathsetmacro{\xs}{0-0.8*\segmentLength}
\pgfmathsetmacro{\xA}{0}
\pgfmathsetmacro{\xf}{0+0.8*\segmentLength}
\threeL{\xs}{\xA}{\y}
\twoL{\xA}{\xf}{\y}
\bub{\xA}{0}{1.5*\y}{#1}{}
\end{feynhand}
\end{tikzpicture}
}

\title{Resumming Scattering Amplitudes for Waveforms
}

\author[,a]{Katsuki Aoki\footnote{Present address: Graduate School of Science and Engineering, Saitama University, 255 Shimo-Okubo, Sakura-ku, Saitama 338-8570, Japan}}
\affiliation[a]{Center for Gravitational Physics and Quantum Information,
Yukawa Institute for Theoretical Physics, Kyoto University, 606-8502, Kyoto, Japan }
\emailAdd{katsukiaoki@mail.saitama-u.ac.jp}
\author[a]{and Andrea Cristofoli} 
\emailAdd{cristofoli@yukawa.kyoto-u.ac.jp}

\abstract{We develop a formalism to compute non-perturbative 5-point scattering amplitudes and apply it to gravitational waveforms in the two-body problem for arbitrary trajectories. Drawing inspiration from Feshbach’s projector formalism in nuclear physics, we introduce effective potentials governing graviton emission and relate them to perturbative scattering amplitudes at arbitrary order in the gravitational coupling and mass ratio. Once these potentials are determined, the corresponding non-perturbative amplitudes in the classical limit are obtained by iterative insertions and subsequently translated into gravitational waveforms using the KMOC formalism. As an application, we compute the gravitational waveform emitted by a conservative two-body dynamics moving along a generic, potentially highly bent, trajectory. Our formalism extends effective field theory matching of the gravitational two-body potential to radiative phenomena, enabling the extraction of gravitational-wave source terms directly from perturbative on-shell amplitudes.
}

\begin{document}
{\baselineskip0pt
\rightline{\baselineskip16pt\rm\vbox to-20pt{
           \hbox{YITP-26-05}
\vss}}%
}

\maketitle

\section{Introduction and summary}
\label{sec:intro}
Following the first decade of gravitational-wave (GW) detections, GW physics has entered an era of precision science~\cite{LIGOScientific:2025slb}. As high-precision waveform modelling cannot rely solely on numerical relativity simulations, analytical inputs are becoming more essential than ever. In response to this need, an old approach to classical physics has been reconsidered, drawing from some of the unexpected corners of modern theoretical physics --- quantum field theory (QFT) and scattering-amplitude methods. The idea of extracting a classical gravitational observables from scattering amplitudes dates back to Iwasaki \cite{Iwasaki:1971iy, Iwasaki:1971vb}, who computed post-Newtonian potentials for a binary from the classical limit of a one-loop amplitude of massive scalars exchanging gravitons, using the conventional Feynman-diagrammatic available at the time. Over the last decades, however, amplitude computations have undergone a dramatic transformation, with on-shell and unitarity-based approaches allowing unprecedented efficiency, without reference to Lagrangians or gauge choices \cite{Bern:1994zx,Bern:1994cg,Bern:1995db,Laporta:2000dsw}. This has established amplitude techniques as some of the most powerful tools in perturbative collider physics, making their extension to gravitational-wave science both natural and essential.

One of the main reasons for the usefulness of such methods for GW physics lies in the Effective One Body (EOB) formalism \cite{Buonanno:1998gg}. 
Ref.~\cite{Damour:2016gwp} proposed that gauge-invariant information extracted from scattering observables in the post-Minkowskian (PM) approximation—a weak-field expansion valid for arbitrary velocities—can be mapped onto and resummed within the EOB framework, thereby providing a powerful resummation of perturbative results applicable to bound systems. This idea was subsequently realized in practice through a remarkable two-loop calculation of 2-to-2 scattering of massive scalars in gravity \cite{Bern:2019crd,Bern:2019nnu}, enabled by an EFT matching approach \cite{Cheung:2018wkq} based on \cite{Neill:2013wsa}. These results were later incorporated into EOB-based waveform models and studied for their impact on improving waveform templates \cite{Antonelli:2019ytb}, leading to an active research field with new results in classical gravity derived from modern amplitude techniques~\cite{Bern:2021dqo, Herrmann:2021lqe, DiVecchia:2021bdo, Herrmann:2021tct, Bjerrum-Bohr:2021vuf, Heissenberg:2021tzo, Bjerrum-Bohr:2021din, Brandhuber:2021eyq, Bern:2021yeh, Bern:2022jvn, DiVecchia:2022nna, DiVecchia:2022piu, Driesse:2024xad, Bern:2024adl, Driesse:2024feo, Heissenberg:2025ocy, Brammer:2025rqo, Bern:2025zno, Bern:2025wyd}.

\begin{figure}[t]
\centering
\includegraphics[width=\linewidth]{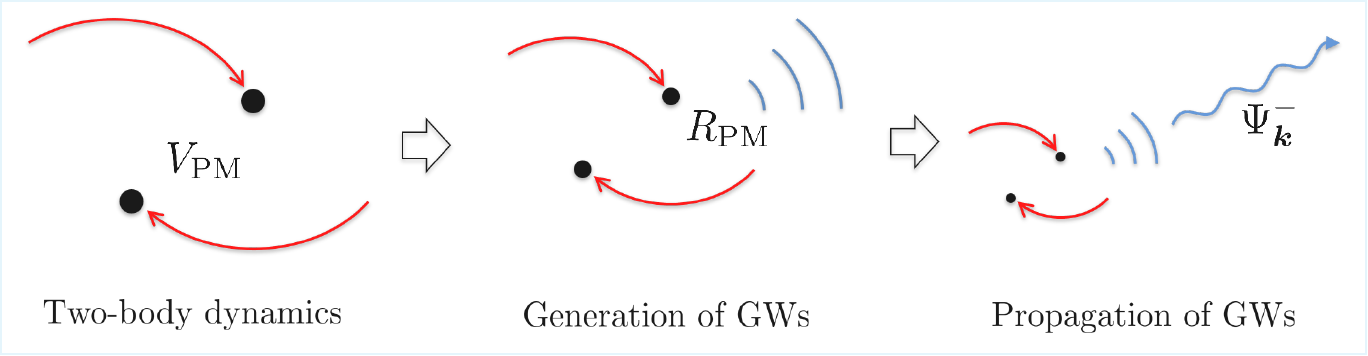}
\caption{The problem of GWs can be split into three stages. In our framework, each is described by the post-Minkowskian two-body potential $V_{\rm PM}$, the radiation potential $R_{\rm PM}$, and the wavefunction of the outgoing graviton $\Psi^-_{\bmk}$. This paper focuses on the generation, with only brief discussions of propagation.}
\label{fig:GWs}
\end{figure}

\begin{figure}[t]
\centering
\begin{minipage}[b]{0.45\linewidth}
\centering
\scalebox{1.5}{
\begin{tikzpicture}
\begin{feynhand}
\def\y{0.5}
\twoL{-2}{1.2}{2*\y}
\node at (0,0) {$\cdots$};
\foreach \x in {1,2,3,4}
\propag[boson] (-0.4*\x, \y)--(-0.4*\x,-\y);
\foreach \x in {1,2}
\propag[boson] (0.4*\x, \y)--(0.4*\x,-\y);
\node at (0,-0.8) {};
\end{feynhand}
\end{tikzpicture}
}
\subcaption{Ladders}
\end{minipage}
\begin{minipage}[b]{0.45\linewidth}
\centering
\scalebox{1.5}{
\begin{tikzpicture}
\begin{feynhand}
\def\y{0.5}
\twoL{-1.8}{1.8}{2*\y}
\propag[boson] (0.1, \y) -- (-1.8, 1);
\foreach \x in {0.5, 1.5, 3.5, -0.5, -2.5, -3.5}
\propag[boson] (0.4*\x, \y)--(0.4*\x,-\y);
\node at (1,0) {$\cdots$};
\node at (-0.6,0) {$\cdots$};
\node at (0,-0.8) {};
\end{feynhand}
\end{tikzpicture}
}
\subcaption{DWBA}
\end{minipage}
\caption{Typical diagrams computed in resummations.}
\label{fig:resummation}
\end{figure}
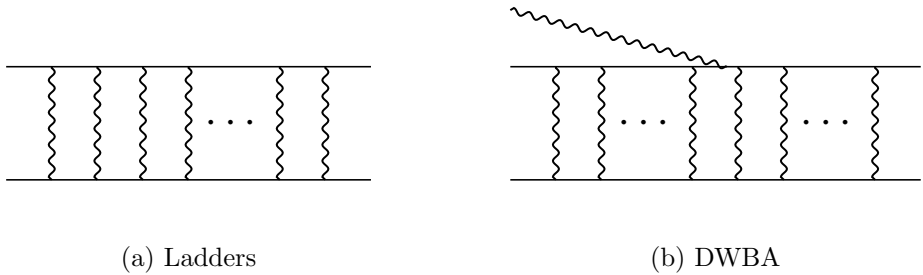

From the perspective of the EOB formalism, although scattering amplitudes provide information about the two-body dynamics, the generation of gravitational waves and their subsequent propagation are inferred separately (see Fig.~\ref{fig:GWs}). This naturally raises a question of whether the amplitude-based framework can continue to improve understanding of these subsequent stages. One answer to this question was provided by the observable-based method, known as the KMOC formalism~\cite{Kosower:2018adc}, which makes it possible to compute gravitational waveforms directly from on-shell scattering amplitudes \cite{ Cristofoli:2021vyo}. Using this but also alternative methods, such as WQFT \cite{Mogull:2020sak,Jakobsen:2021smu}
and eikonal \cite{DiVecchia:2023frv} for example, several state-of-the-art results for scattering waveforms have been derived~\cite{Brandhuber:2023hhy, Herderschee:2023fxh, Georgoudis:2023lgf, Caron-Huot:2023vxl, Alessio:2024wmz, Alessio:2024onn, Fucito:2024wlg, Georgoudis:2025vkk, Brunello:2025eso, Heissenberg:2025fcr}. The main challenge, however, is that powerful amplitude methods such as generalized unitarity are currently available only in the perturbative regime. The perturbative expansion in $G$ fails for small relative velocities, even for scattering in the weak-gravity region, where the classical trajectories of the particles are highly bent, i.e., correspond to a large scattering angle. Predicting bound states (bound orbits) also requires a careful resummation of iterated topologies which to date is not fully understood (the left panel of Fig.~\ref{fig:resummation}). These difficulties highlight the complexity in applying amplitude techniques to calculating waveforms valid in a wide range of astronomical situations.

A variety of attempts have recently been made to deepen our understanding of scattering amplitudes beyond perturbation theory. Connecting with a common theme that classical physics can be recovered from exponentiated quantities \cite{Damgaard:2021ipf,Damgaard:2023ttc}, a proposal was given in \cite{Cristofoli:2021jas} 
focused on the semiclassical properties of the final state of a scattering process and the KMOC formalism. In \cite{Cristofoli:2021jas}, the resummation of elastic amplitudes was understood as a result of eikonal exponentiation, while the resummation of inelastic amplitudes was viewed as arising from the coherence of the final state describing gravitational radiation.\footnote{However, the assumption that gravitational radiation emitted by a two-body system is in a coherent state does not hold generally. As pointed out in~\cite{Fernandes:2024xqr,Fernandes:2025cog,Georgoudis:2025vkk,Kanno:2025how}, a squeezed-state description may be more appropriate.} For other attempts related to eikonal and radiation physics, see
\cite{Kim:2025hpn, Alessio:2025flu, Kim:2025olv, Kim:2025gis, Akhtar:2025fil}. However, our work does not take such exponentiations of amplitudes as its starting point. Instead, returning to older ideas—mainly developed in the context of nuclear physics—and combining them with modern developments in scattering amplitudes and gravitational waves, we address the problem of resumming the perturbation series of scattering amplitudes for radiation.

One of the ideas is the so-called distorted-wave Born approximation (DWBA), which was developed in the context of quantum scattering theory, particularly nuclear reaction physics. The DWBA replaces plane waves with exact solutions of a certain part of the Hamiltonian, such as Coulomb wavefunctions in nonrelativistic QED. It can be applied, for example, to calculate radiative transition rates of atoms or molecules as shown in Chapter 11 of~\cite{weinberg2015lectures}. From a quantum field theory perspective, the use of Coulomb wavefunctions can be understood as resumming iterated photon exchanges in the nonrelativistic regime, both before and after radiation emission (see the right panel of Fig.~\ref{fig:resummation}). The recent use of background-based approaches to understand the self-force expansion can be regarded as a version of the same approximation scheme where we replace plane wave states with solutions to the wave equation in a curved spacetime. Using the Schwarzschild spacetime as a background, one can then compute geodesic motion \cite{Kol:2021jjc,Adamo:2022rmp,Adamo:2021rfq}, one-graviton emission amplitudes for a two-body scattering \cite{Adamo:2023cfp,Cristofoli:2025esy}, waveforms \cite{Khalaf:2023ozy, Khalaf:2025jpt} or merger~\cite{Aoki:2024boe} to all orders in the gravitational coupling. They can indeed calculate non-perturbative amplitudes, but are applicable only in certain situations: the DWBA for non-relativistic systems and the background approach for the extreme mass ratio limit.

\begin{figure}[t]
\centering
\begin{minipage}[b]{0.9\linewidth}
\begin{align*}
\begin{tikzpicture}[baseline=-2]
\begin{feynhand}
\def\y{0.5}
\twoL{-1.3}{1.3}{2*\y}
\bub{0}{0}{1.5*\y}{V_{\rm PM}}{}
\end{feynhand}
\end{tikzpicture}
~&=~
\begin{tikzpicture}[baseline=-2]
\begin{feynhand}
\def\y{0.5}
\twoL{-1}{1}{2*\y}
\propag[boson] (0,\y) -- (0,-\y); 
\end{feynhand}
\end{tikzpicture}
~+~
\begin{tikzpicture}[baseline=-2]
\begin{feynhand}
\def\y{0.5}
\twoL{-1}{1}{2*\y}
\propag[boson] (0.5,\y) -- (0,-\y) -- (-0.5,\y); 
\end{feynhand}
\end{tikzpicture}
~+~
\begin{tikzpicture}[baseline=-2]
\begin{feynhand}
\def\y{0.5}
\twoL{-1}{1}{2*\y}
\propag[boson] (0.5,-\y) -- (0,\y) -- (-0.5,-\y); 
\end{feynhand}
\end{tikzpicture}
~+~ \cdots
\\
\begin{tikzpicture}[baseline=-2]
\begin{feynhand}
\def\y{0.5}
\twoL{0}{1.3}{1.5*\y}
\threeL{-1.3}{0}{\y}
\bub{0}{0}{1.5*\y}{R_{\rm PM}}{}
\end{feynhand}
\end{tikzpicture}
~&=~
\begin{tikzpicture}[baseline=-2]
\begin{feynhand}
\def\y{0.5}
\twoL{-1}{1}{1.5*\y}
\propag[boson] (0,0.75*\y) -- (-1, 1.5*\y);
\end{feynhand}
\end{tikzpicture}
~+~
\begin{tikzpicture}[baseline=-2]
\begin{feynhand}
\def\y{0.5}
\twoL{-1}{1}{1.5*\y}
\propag[boson] (0, -0.75*\y) -- (-1, -1.5*\y);
\end{feynhand}
\end{tikzpicture}
~+~
\begin{tikzpicture}[baseline=-2]
\begin{feynhand}
\def\y{0.5}
\twoL{-1}{1}{2*\y}
\propag[boson] (0,\y) -- (0, -\y);
\propag[boson] (0,0) -- (-1, 0);
\end{feynhand}
\end{tikzpicture}
~+~ \cdots
\end{align*}
\subcaption{Typical diagrams in effective potentials.}
\centering
\begin{minipage}[b]{0.9\linewidth}
\begin{align*}
    \begin{tikzpicture}[baseline=-2]
\begin{feynhand}
\def\y{0.5}
\twoL{-5.5}{5.5}{2*\y}
\propag[boson] (0, 1.2*\y) -- (-5.5, 3*\y);
\bub{0}{0}{1.5*\y}{R_{\rm PM}}{}
\foreach \x in {1, 2.5, -1, -2.5}
\bub{1.7*\x}{0}{1.5*\y}{V_{\rm PM}}{};
\node at (1.7*1.75,0) {$\cdots$};
\node at (-1.7*1.75,0) {$\cdots$};
\draw [
    thick,
    decoration={
        brace,
        mirror
    },
    decorate
] (1, -0.9) -- (5,-0.9) ;
\node at (3,-1.4) {$\ket{\Psi^+_{\bmp}}$};
\draw [
    thick,
    decoration={
        brace
    },
    decorate
] (-1, -0.9) -- (-5,-0.9) ;
\node at (-3,-1.4) {$\bra{\Psi^-_{\bmp'}}$};
\end{feynhand}
\end{tikzpicture}
\end{align*}
\subcaption{Iterated topologies are resummed by using wavefunctions.}
\end{minipage}
\end{minipage}
\caption{Schematic structure of our resummed amplitudes.}
\label{fig:resum_amp}
\end{figure}
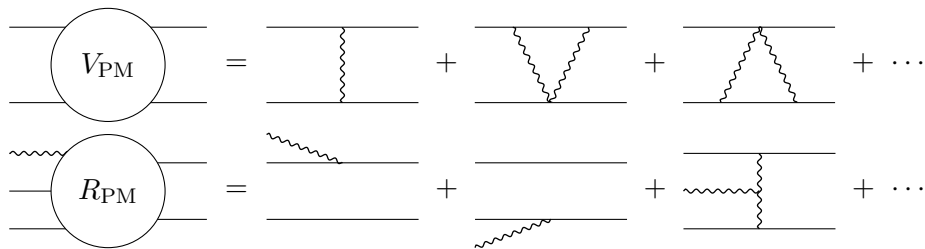
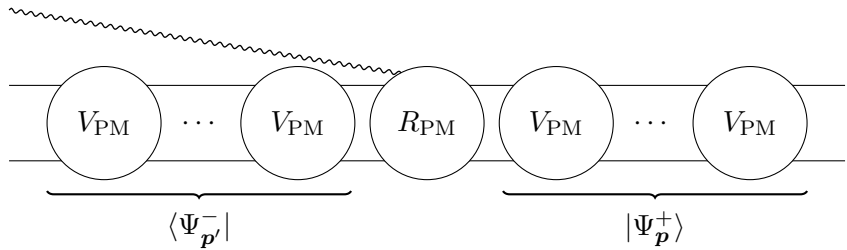

The key idea --- especially for extending these nonrelativistic and background-based approaches to relativistic scattering theory with generic mass ratios --- is the projector formalism of Feshbach~\cite{Feshbach:1958nx,Feshbach:1962ut} (see \cite{Muga:2004zz} for a review). The Feshbach formalism can be regarded as a precursor of effective theory, as it defines non-Hermitian effective potentials from an underlying Hamiltonian via projection operators (see the top panel of Fig.~\ref{fig:resum_amp} for an illustration). This approach is similar to, but distinct from, effective field theory: in the former, certain states are projected out, whereas in the latter, certain fields are integrated out.
For example, within the Feshbach formalism, one may project out states containing multiple gravitons while retaining states with a single graviton; by contrast, integrating out the graviton field renders all graviton states invisible. An important observation is that the effective potential obtained by applying the Feshbach formalism to the two-body system, i.e., retaining only two-massive-particle states, agrees with the two-body potential derived from EFT approaches~\cite{Neill:2013wsa,Cheung:2018wkq} or more traditional Lippmann-Schwinger methods \cite{Iwasaki:1971iy, Iwasaki:1971vb, Cristofoli:2019neg}. The advantage of adopting the Feshbach formalism is that the notion of effective potentials is not restricted to conservative systems and can be naturally extended to radiative processes by retaining graviton states as well. At the same time, we preserve the benefits of modern amplitude-based methods, in which effective potentials are directly extracted from perturbatively computed scattering amplitudes, by employing the method of Born subtraction~\cite{Cristofoli:2019neg} (see below).

The essential ingredient of this paper is the {\it radiation potential} $R_{\rm PM}$ responsible for graviton emission processes. We will show that the emission amplitude is computed by iterating the two-body potential $V_{\rm PM}$ before and after the radiation potential, as illustrated in the bottom panel of Fig.~\ref{fig:resum_amp}. As in the DWBA mentioned above, the iterated diagrams can then be resummed by replacing the plane waves with solutions to the effective Schr\"{o}dinger equations.\footnote{A similar workflow is used in~\cite{Iengo:2009ni} in the context of Sommerfeld enhancement, which we extend to generic amplitudes through the Feshbach projection and the Born subtraction.} The resummed scattering amplitude is then used to compute the waveform via the KMOC formalism~\cite{Kosower:2018adc, Cristofoli:2021vyo}. We will derive the generic formulae of the resummed amplitude and waveform under generic forms of the effective potentials. They are thus valid for all orders in $G$ and indeed exhibit the same structure as the amplitudes in the background approach, but for generic mass ratios. It will turn out that the generation of gravitational waves from classical conservative dynamics can be computed by simply evaluating the radiation potential along classical trajectories. This provides a shortcut for mapping perturbative amplitudes to non-perturbative waveforms. As a proof of concept, we explicitly compute the radiation potential $R_{\rm PM}$
at leading order in $G$ and show that the KMOC waveform exactly agrees with the gravitational waves emitted from a generic trajectory of two worldlines, including highly bent trajectories.
\\

\noindent
{\bf Summary of results.} We summarise our main results below for ease of reference:\begin{itemize}
\item {\bf Effective potentials.} Suppose we have calculated scattering amplitudes at a given order in the PM expansion. These perturbative inputs are mapped into the two-body potential $V_{\rm PM}$ and the radiation potential $R_{\rm PM}$ at that order $G$ by one-channel and two-channel Born subtractions, respectively. In a diagrammatic notation explained in Sec.~\ref{sec:example}, the formulae are 
\begin{align}
\Atwo{V_{\rm PM}} \, &=\, \Atwo{T} - 
\begin{tikzpicture}[baseline=-2]
\begin{feynhand}
\twoL{-1.4}{1.34}{0.3}
\bub{-0.6}{0}{1.5*0.3}{T}{}
\bub{0.6}{0}{1.5*0.3}{T}{}
\end{feynhand}
\end{tikzpicture}
+ \cdots
\,, \\
    \Arad{R_{\rm PM}}\, &= \,\Arad{T}
    -
\begin{tikzpicture}[baseline=-2]
\begin{feynhand}
\def\y{0.3}
\def\segmentLength{1}
\pgfmathsetmacro{\xs}{0-0.8*\segmentLength}
\pgfmathsetmacro{\xA}{0}
\pgfmathsetmacro{\xB}{0+\segmentLength}
\pgfmathsetmacro{\xf}{0+\segmentLength+0.65*\segmentLength}
\threeL{\xs}{\xA}{\y}
\twoL{\xA}{\xf}{\y}
\bub{\xA}{0}{1.5*\y}{T}{}
\bub{\xB}{0}{\y}{T}{}
\end{feynhand}
\end{tikzpicture}
-
\begin{tikzpicture}[baseline=-2]
\begin{feynhand}
\def\y{0.3}
\def\segmentLength{1}
\pgfmathsetmacro{\xs}{0-\segmentLength}
\pgfmathsetmacro{\xA}{0-0.2*\segmentLength}
\pgfmathsetmacro{\xB}{0+\segmentLength}
\pgfmathsetmacro{\xf}{0+\segmentLength+0.8*\segmentLength}
\threeL{\xs}{\xB}{\y}
\twoL{\xB}{\xf}{\y}
\bub{\xA}{0}{1.5*\y}{T}{}
\bub{\xB}{0}{1.5*\y}{T}{}
\end{feynhand}
\end{tikzpicture}
+\cdots
\,.
\end{align}
Here, $T$ denotes the $T$-matrix whose connected parts correspond to an off-energy-shell extension of the scattering amplitudes. Note that one can also derive the effective potential for propagating GWs. However, our main focus is the generation of the GWs, so we will not discuss the propagation in detail in this paper.

\item {\bf Scattering and inclusive amplitudes.} The resummed 5-point scattering amplitude and inclusive amplitude are computed by sandwiching the radiation potential between the wavefunctions:
\begin{alignat}{2}
    \braket{{p_{1};p_{2};k|\hat{S}|p_{1};p_{2}}}
    &\,\eqcl &&-i\sqrt{32E_{1} E_{2} E'_{1} E'_{2} \omega}\, \hdelta^{(4)}(P'+k-P) \braket{\Psi^-_{\bmp'};\Psi^-_{\bmk}|\hat{R}_{\rm PM}|\Psi^+_{\bmp}}
    \,, \\
     \braket{p'_{1};p'_{2}|\hat{S}^{\dagger}\hat{a}_{k}\hat{S}|p_{1};p_{2}}
    &\eqcc &&-i\sqrt{32E_{1} E_{2} E'_{1} E'_{2} \omega}\, \hdelta^{(4)}(P'+k-P) \braket{\Psi^+_{\bmp'};\Psi^-_{\bmk}|\hat{R}_{\rm PM}|\Psi^+_{\bmp}}
    \,,
\end{alignat}
where $p_a^{\mu}=(E_a,\bmp_a)$ with $a=1,2$ and $k^{\mu}=(\omega,\bmk)$ are the on-shell momenta for the massive particle $a$ and the graviton; $P$ and $p$ are the total and relative momenta of massive particles; $\Psi^{\pm}_{\bmp}$ are the in/out wavefunctions of massive bodies whereas $\Psi^-_{\bmk}$ is that of outgoing graviton; and the ``BO'' and ``cons'' denote equalities under the Born-Oppenheimer-type approximation and a certain assumption on the conservative dynamics described in \eqref{three-body_state} and \eqref{completness_cons}, respectively. As we mentioned above, we only give a brief discussion on the propagation, so the graviton wavefunction is undetermined in this paper (or is assumed to be free propagation $\Psi^-_{\bmk}(\bm{y})=e^{i\bmk\cdot \bm{y}}$). Our concern is the wavefunction of the massive bodies, which we solve for a generic two-body potential under the WKB approximation. The solution takes the form
\begin{align}
    \Psi^+_{\bmp}(\bm{x})=\sqrt{\det \partial_{\bmp} \partial_{\bmx} \sigma_{\bmp}(\bmx)}\,e^{i\sigma_{\bmp}(\bm{x})}
    \,,
\end{align}
with $\sigma_{\bmp}$ satisfying \eqref{eq_bmp}.

\item {\bf Waveform.} As shown in Fig.~\ref{fig:GWs}, the calculation of waveform consists of three stages. Working in the centre-of-mass (COM) frame of the two bodies, we will show that the resummed waveform computed by the KMOC formalism indeed exhibits such a structure; concretely, the waveform $W(k)$ is given by
\begin{align}
W&\eqcc \int \dd^3 \bm{y}  [\Psi^-_{\bmk}(\bm{y})]^* \mathcal{T}(\bm{y})\,, \quad
    \mathcal{T}(\bm{y})=\int \hd^3 \bml e^{-i\bml\cdot \bm{y}}\,\mathcal{T}(\bml)
    \label{W_intro}
    \,, \\
    \mathcal{T}(\bml)&~=-\sqrt{2\omega}\int \dd t \  e^{i\omega t} 
     e^{-i\bmX_{\rm cl}(t)\cdot \bmk}
     R_{\rm PM}(\bmp_{\rm cl}(t),\bml,\bmx_{\rm cl}(t))
    \,,
    \label{Tk_intro}
\end{align}
where $\bmx_{\rm cl}(t), \bmp_{\rm cl}(t)$ and $\bmX_{\rm cl}(t)$ are the relative position, the relative momentum, and the COM position of the classical two bodies at time $t$. Hence, the computation of the waveform is reduced to integrating the radiation potential along the classical trajectory under the two-body potential and then performing the overlap integral of the graviton wavefunction $\Psi^-_{\bmk}$ and the source term $\mathcal{T}(\bm{y})$. For free propagating GWs, the spectral waveform is directly given by the source term $W=\mathcal{T}(\bmk)$. Note that we will derive these formulae purely based on the scattering amplitudes. We do not a priori assume classical equations of motion or bulk spacetime; these notions emerge from the classical limit of the resummed scattering amplitudes.
\end{itemize}

\noindent
{\bf Outline.} The rest of the paper is organised as follows. In Sec.~\ref{sec:non-pert}, starting with a brief review of the Born series and the DWBA, we introduce the Feshbach formalism and show that the exact $T$-matrix can be written in terms of iterative insertions of effective potentials. The results of this section are general and applicable to arbitrary inelastic amplitudes. We then specialise our formalism to the 2-to-3 amplitude in Sec.~\ref{sec:resum}. There, we elaborate on the procedures of (i) mapping perturbative amplitudes to effective potentials, (ii) constructing non-perturbative wavefunctions under the WKB approximation, and (iii) using them to compute the non-perturbative 5-point amplitude, in that order. The analysis in Sec.~\ref{sec:resum} relies on the classical limit and a factorisation assumption for the three-particle state motivated by the large hierarchy between massive and massless momenta, but does not require Hermiticity of the effective potentials. In Sec.~\ref{sec:waveform}, we compute waveforms under Hermitian effective potentials, corresponding to radiation emitted from conservative dynamics. We show that the notion of classical bulk physics emerges naturally from the resummed amplitudes and the KMOC formalism, and we derive the waveform formulae \eqref{W_intro} and \eqref{Tk_intro} for generic Hermitian potentials. We conclude and discuss future directions in Sec.~\ref{sec:conclusion}. Appendix~\ref{sec:saddle} provides a general formula for a multivariable saddle-point integral up to subleading order and describes the WKB approximation of the relativistic Schr\"{o}dinger equation. We also calculate the impulse for conservative dynamics in Appendix~\ref{sec:impulse}, showing that our non-perturbative wavefunction reproduces the classical result.
\\

\noindent
{\bf Notations.} The metric signature is $\eta_{\mu\nu}={\rm diag}[+1,-1,-1,-1]$. Quantities with hats are operators, except for the integration measure and the delta function with $2\pi$:
\begin{align}
\hd p=\frac{\dd p}{2\pi}\,, \qquad \hdelta(p)=2\pi\delta(p)
\,.
\end{align}
The Lorentz-invariant phase-space measure is denoted by
\begin{align}
d \Phi(p_1,p_2,\cdots) = \prod_a \hd^4 p_a \hdelta(p_a^2-m_a^2) \theta(p^0_a) = \prod_a \frac{\hd^3 \bm{p}_a}{2E_a} 
\end{align}
and an on-shell one-particle state is normalised as
\begin{align}
    \braket{p_1|p'_1}=2E_1\hdelta^{(3)}(\bm{p}_1-\bm{p}'_1)
    \,,
\end{align}
with $E_1=\sqrt{m_1^2+\bm{p}_1^2}$. Bold characters like $\bm{p},\bm{x}$, etc.~are spatial vectors unless otherwise stated. The free states are denoted by using their labels, e.g., $\ket{p_1}, \ket{p_1;p_2}=\ket{p_1}\otimes \ket{p_2}$, whereas interacting states are represented by $\ket{\Psi^{\pm}_{\alpha}}$ and $\ket{\psi^{\pm}_{\alpha}}$ with the surperscript denoting the boundary condition ($+$: retarded, $-$: advanced) and the subscript $\alpha$ being the labels of the state.


\section{Reorganisations of Born series}
\label{sec:non-pert}
\subsection{Born series}
Let us start with the Lippmann-Schwinger equation (see, e.g.~\cite{weinberg2015lectures, Weinberg:1995mt})
\begin{align}
\ket{\Psi_{\alpha}^{\pm}} &=\ket{\alpha} + \frac{1}{E_{\alpha}-\hat{H}_0 \pm i \varepsilon} \hat{V} \ket{\Psi_{\alpha}^{\pm}}
\label{LPeq}
\end{align}
where the Hamiltonian $\hat{H}$ is assumed to split into the free and interaction parts $\hat{H}=\hat{H}_0+\hat{V}$. The in/out and free states, which can be generic multi-particle states, are the eigenstates of the exact and free Hamiltonians:
\begin{align}
\hat{H}\ket{\Psi_{\alpha}^{\pm}}=E_{\alpha}\ket{\Psi^{\pm}_{\alpha}}\,, \qquad \hat{H}_0 \ket{\alpha}=E_{\alpha}\ket{\alpha}
\,,
\end{align}
where $\alpha$ is the collective index for a state. The $S$-matrix is given by
\begin{align}
S_{\beta,\alpha}=\braket{\Psi_{\beta}^- | \Psi_{\alpha}^+} &=\braket{\beta|\alpha} - i \hdelta(E_{\alpha}-E_{\beta}) T_{\beta,\alpha}
\,,
\label{S-matrix}
\end{align}
with the $T$-matrix
\begin{align}
T_{\beta,\alpha} &:= \bra{\beta} \hat{V} \ket{\Psi^{+}_{\alpha} }
\,.
\label{T-matrix}
\end{align}
Note that the $T$-matrix is defined even for $E_{\alpha}\neq E_{\beta}$ (whereas each particle satisfies the on-mass-shell condition $p^2=m^2$). The off-energy-shell $T$-matrix cannot be unique, and this ambiguity is related to the freedom of the Hamiltonian under canonical transformations.

By iteratively solving the Lippmann-Schwinger equation \eqref{LPeq}\footnote{Precisely speaking, the Lippmann-Schwinger equation for multi-particle states cannot give a unique solution, which can be seen as a divergence due to delta functions of disconnected diagrams in perturbation theory. While there are ways to deal with this problem~\cite{Faddeev:1960su, Weinberg:1964zza}, we accept the Born series as the formal solution to \eqref{LPeq} because these subtleties do not enter, at least in the analysis of this paper.}
\begin{align}
    \ket{\Psi^+_{\alpha}}=\sum_{n=0}^{\infty} (\hat{G}_0^+ \hat{V})^n\ket{\alpha} = \ket{\alpha}+\hat{G}_0^+ \hat{V} \ket{\alpha}+\cdots
    \,,
\end{align}
we obtain the Born series of the $T$-matrix
\begin{align}
    T_{\beta,\alpha}
    &=\sum_{n=0}^{\infty} \bra{\beta}\hat{V}(\hat{G}_0^+ \hat{V})^n\ket{\alpha}
    \nn
    &=\braket{\beta|\hat{V}|\alpha}+\braket{\beta|\hat{V}\hat{G}_0^+\hat{V}|\alpha}+\cdots
    \,, \label{Born_series}
\end{align}
where
\begin{align}
    \hat{G}^+_0:=\frac{1}{E_{\alpha}-\hat{H}_0+i\varepsilon}
    \,.
\end{align}
The Born series \eqref{Born_series} is understood to be the perturbation theory in $\hat{V}$ where all interactions are treated equally. On the other hand, we are interested in resumming the Born series by focusing on iterated topologies as shown in Fig~\ref{fig:resum_amp}. This requires reorganising the Born series into appropriate forms, which we develop in the following subsections.

\subsection{Distorted-wave Born series}
A well-known non-perturbative method is the distorted-wave Born approximation (DWBA), see \cite{weinberg2015lectures} for instance. Let us first split the interactions into the ``strong" and ``weak" parts
\begin{align}
    \hat{V}=\hat{V}_s+\hat{V}_w\,,
\end{align}
and define the ``strong'' Hamiltonian
\begin{align}
   \hat{H}_s:=\hat{H}_0 + \hat{V}_s
    \,.
\end{align}
The strong part $\hat{V}_s$ is what we want to keep non-perturbatively.

We introduce the states $\ket{\Psi^{\pm}_{s,\alpha}}$ as the solutions of the strong Hamiltonian
\begin{align}
    \hat{H}_s\ket{\Psi^{\pm}_{s,\alpha}}=E_{\alpha}\ket{\Psi^{\pm}_{s,\alpha}}
    \,, \label{strong_Sch}
\end{align}
which can be formally solved by the Lippmann-Schwinger equation
\begin{align}
    \ket{\Psi^{\pm}_{s,\alpha}}=\ket{\alpha}+\frac{1}{E_{\alpha}-H_0 \pm i \varepsilon}\hat{V}_s \ket{\Psi^{\pm}_{s,\alpha}}\,.
    \label{LPs}
\end{align}
They are the in/out states if $\hat{V}_w$ were zero. Then, the $T$-matrix can be written as
\begin{align}
    T_{\beta,\alpha}
    &=\braket{\beta|\hat{V}|\Psi^+_{\alpha}}
    \nn
    &=\braket{\Psi_{s,\beta}^-|\hat{V}|\Psi^+_{\alpha}}-\braket{\Psi^-_{s,\beta}|\hat{V}_s \hat{G}_0^+ \hat{V}|\Psi^+_{\alpha}}
    \nn
    &= \braket{\Psi^-_{s,\beta}|\hat{V}_s|\alpha} + \braket{\Psi^-_{s,\beta}|\hat{V}_w|\Psi^+_{\alpha}} 
\end{align}
where we have used \eqref{LPs} to get the second line and then used \eqref{LPeq} and $\hat{V}=\hat{V}_s+\hat{V}_w$ to reach the last line. Then, we express the exact state $\ket{\Psi^+_{\alpha}}$ by using the strong state $\ket{\Psi^+_{s, \alpha}}$. This can be achieved by rewriting the exact Schr\"{o}dinger equation \eqref{LPeq} in the form
\begin{align}
(E_{\alpha}-\hat{H}_s)\ket{\Psi^{\pm}_{\alpha}}=\hat{V}_w \ket{\Psi^{\pm}_{\alpha}}
\,,
\end{align}
and solving it under the boundary condition: $\ket{\Psi^{\pm}_{\alpha}} \to \ket{\Psi^{\pm}_{s,\alpha}}$ as $\hat{V}_w \to 0$. The solution is
\begin{align}
    \ket{\Psi^{\pm}_{\alpha}}=\ket{\Psi^{\pm}_{s,\alpha}}+\hat{G}_s^{\pm}\hat{V}_w \ket{\Psi^{\pm}_{\alpha}} = \sum_{n=0}^{\infty} [\hat{G}_s^{\pm}\hat{V}_w ]^n\ket{\Psi^{\pm}_{\alpha}}
\end{align}
with
\begin{align}
    \hat{G}_s^{\pm} = \frac{1}{E_{\alpha}-\hat{H}_s \pm i \varepsilon}
    \,.
\end{align}
As a result, we obtain the following form of the $T$-matrix
\begin{align}
    T_{\beta,\alpha}=\braket{\Psi^-_{s,\beta}|\hat{V}_s|\alpha} + \sum_{n=0}^{\infty} \braket{\Psi^-_{s,\beta}|\hat{V}_w(\hat{G}_s^{\pm}\hat{V}_w )^n|\Psi^+_{s,\alpha}} 
    \,. \label{DWB_series}
\end{align}
The distorted-wave Born series \eqref{DWB_series} is the series expansion valid only in the weak interaction $\hat{V}_w$. Therefore, if we can solve the ``strong'' Schr\"{o}dinger equation \eqref{strong_Sch}, we can retain the effects of the strong potential to all orders. In particular, when the process $\alpha \to \beta$ cannot take place in the absence of $\hat{V}_w$, the leading order approximation of the $T$-matrix is
\begin{align}
    T_{\beta,\alpha}\simeq \braket{\Psi^-_{s,\beta}|\hat{V}_w|\Psi^+_{s,\alpha}} 
    \,,
    \label{LDWB}
\end{align}
which is linear in the weak interaction $\hat{V}_w$ but non-perturbative in the strong interaction $\hat{V}_s$.

For instance, one can apply the DWBA to non-relativistic QED in order to compute a radiation emission from bound orbits (bound state). In the non-relativistic limit, we can interpret the time and spatial components of the Maxwell field as the ``potential'' and ``radiation'' modes, respectively. Since the time component is non-dynamical, it can be integrated out to yield the Hamiltonian $\hat{H}=\hat{H}_s + \hat{V}_w$ with
\begin{align}
    \hat{H}_s&=\hat{H}_{0,\gamma}+\sum_{a=1,2}\frac{\hat{\bm{p}}_a^2}{2m_a}+\Phi_{\rm Coulomb}(|\bm{x}_1-\bm{x}_2|)
    \,, \label{QED_strong} \\
    \hat{V}_w&=\sum_{a=1,2}\left[ -\frac{e_a}{m_a}\hat{\bm{A}}(\bm{x}_a)\cdot \hat{\bm{p}}_a+\frac{e_a^2}{2m_a}\hat{\bm{A}}^2(\bm{x}_a)\right]
    \,, \label{QED_weak}
\end{align}
where $\hat{\bm{A}}$ is the photon field in the Coulomb gauge, $\hat{H}_{0,\gamma}$ is the free Hamiltonian of the photon, and $m_a$ and $e_a$ are masses and charges of particle $a=1,2$. The Coulomb potential $\Phi_{\rm Coulomb}$, which arises after integrating out the time component of the Maxwell field, is included in the ``strong'' Hamiltonian. One can see that the ``weak'' interaction $\frac{e_a}{m_a}\hat{\bm{A}}\cdot \hat{\bm{p}}_a$ (yielding the Lorentz force and bremsstrahlung) is suppressed by the velocity, while there is no such a suppression in the ``strong'' interaction $\Phi_{\rm Coulomb}$. Therefore, the ``strong'' and ``weak'' interactions are clearly separated in small velocities $\bm{v}_a\ll 1$. Indeed, we can solve the Schr\"{o}dinger equation for the ``strong'' Hamiltonian using the Coulomb wavefunction. The DWBA can then compute the photon-emission amplitude from a non-relativistic bound state trapped by the Coulomb potential~\cite{weinberg2015lectures}.

However, it would not be straightforward to add relativistic corrections to the DWBA. Since the exchange of a photon via $\frac{e_a}{m_a}\hat{\bm{A}}\cdot \hat{\bm{p}}_a$ generates the Lorentz force, one would wish to iteratively insert these photon exchanges to discuss relativistic corrections to dynamics.\footnote{This type of issue arises when solving a differential equation perturbatively, and can be addressed by the renormalisation-group method (see e.g.,~\cite{Kunihiro:1995zt}).} Eventually, we lose the separation between the ``strong'' and ``weak'' interactions in a fully relativistic system $\bm{v}_a=\mathcal{O}(1)$. These difficulties trace back to the fact that the distinction between ``potential'' and ``radiation'' modes of photons becomes subtle in relativistic systems. The same issue persists and becomes even more complicated in gravity due to self-interactions. This observation demands us to develop an alternative method to compute non-perturbative scattering amplitudes without splitting the field into the ``potential'' and ``radiation'' modes, or even without relying on the notion of fields.

\subsection{Feshbach projection formalism}
\label{sec:Feshbach}
We provide another expression of the $T$-matrix by applying the projector theory of Feshbach~\cite{Feshbach:1958nx,Feshbach:1962ut}. Let $\hat{P}$ be the projector selecting the channels of our interest and $\hat{Q}=\hat{1}-\hat{P}$ be the projector for others. Specifically, we choose the initial and final channels as the $P$ state\footnote{The phase-space integral is implicit in this notation. When $\alpha$ denotes the distinguishable two-particle state, it should read $\ket{\alpha}\bra{\alpha}=\int \dd\Phi(p_1,p_2)\ket{p_1;p_2}\bra{p_1;p_2}$.}
\begin{align}
    \hat{P}=\ket{\alpha}\bra{\alpha} + \ket{\beta}\bra{\beta}
    \,.
\end{align}
Hereafter, $\alpha$ and $\beta$ denote certain specific states: for example, $\alpha$ is a two-particle state and $\beta$ is a three-particle state, respectively.
By using the projection operators, the Schr\"{o}dinger equation $(E_{\alpha}-\hat{H})\ket{\Psi^{\pm}_{\alpha}}=0$ can be split into two components
\begin{align}
    (E_{\alpha}-\hat{H}_0-\hat{V}_{P,P})\ket{\hat{P}\Psi^{\pm}_{\alpha}} - \hat{V}_{P,Q}\ket{\hat{Q}\Psi^{\pm}_{\alpha}} &=0
    \,, \label{SchP} \\
    -\hat{V}_{Q,P}\ket{\hat{P} \Psi^{\pm}_{\alpha}}+(E_{\alpha}-\hat{H}_0-\hat{V}_{Q,Q}) \ket{\hat{Q}\Psi^{\pm}_{\alpha}}&=0
    \,, \label{SchQ}
\end{align}
where $\hat{V}_{P,P}=\hat{P}\hat{V}\hat{P}, \ket{\hat{P}\Psi^{\pm}_{\alpha}}=\hat{P}\ket{\Psi^{\pm}_{\alpha}}$ and others are similarly defined.
The formal solutions are
\begin{align}
    \ket{\hat{Q}\Psi_{\alpha}^{\pm}} = \frac{1}{E_{\alpha}-\hat{H}_0-\hat{V}_{Q,Q}\pm i \varepsilon} \hat{V}_{Q,P} \ket{\hat{P} \Psi^{\pm}_{\alpha}}
    \,.
\end{align}
Then, substituting it into \eqref{SchP}, we obtain the two-channel effective Schr\"{o}dinger equation
\begin{align}
    (E_{\alpha}-\hat{H}_0 - \hat{W}_{P,P}^{\pm})\ket{\hat{P}\Psi^{\pm}_{\alpha}}=0\,, 
    \label{couple_Sch}
\end{align}
with the complex potential
\begin{align}
    \hat{W}^{\pm}_{P,P} =\hat{V}_{P,P}+ \hat{V}_{P,Q}\frac{1}{E_{\alpha}-\hat{H}_0-\hat{V}_{Q,Q}\pm i\varepsilon}\hat{V}_{Q,P}
    \,.
    \label{WPP}
\end{align}
The $\pm$ potentials are related via the Hermitian conjugate $\hat{W}^-_{P,P}=(\hat{W}^+_{P,P})^{\dagger}$. The $T$-matrix for the states of interest can then be written as
\begin{align}
    T_{\beta,\alpha}&=\braket{\beta|\hat{V}|\Psi^+_{\alpha}}
    \nn
&=\braket{\beta|\hat{V}_{P,P}|\hat{P}\Psi^+_{\alpha}}+\braket{\beta|\hat{V}_{P,Q}|\hat{Q}\Psi^+_{\alpha}}
\nn
&=\braket{\beta|\hat{W}^+_{P,P}|\hat{P}\Psi^+_{\alpha}}
\,.
\label{T_F1}
\end{align}
Hence, the $T$-matrix is given by a similar form of the original expression \eqref{T-matrix}, but as a (generically) non-Hermitian system with the two-channel complex potential \eqref{WPP}.

Our idea is to take the Feshbach projection once again to further rewrite the $T$-matrix into a form close to the DWBA. The state $\ket{\hat{P}\Psi^+_{\alpha}}=\ket{\alpha}\braket{\alpha|\Psi^+_{\alpha}}+\ket{\beta}\braket{\beta|\Psi^+_{\alpha}}$ contains two different final channels. We introduce the projection operators
\begin{align}
    \hat{\alpha}=\ket{\alpha}\bra{\alpha}\,, \qquad \hat{\beta}=\ket{\beta}\bra{\beta}\,, \qquad 
    \hat{\mathcal{Q}}=\hat{\beta}+\hat{Q}
    \,,
\end{align}
which satisfy $\hat{1}=\hat{\alpha}+\hat{\mathcal{Q}}$. Applying the same argument as \eqref{couple_Sch} and \eqref{WPP} with the replacements $\hat{P}\to \hat{\alpha}, \hat{Q}\to\hat{\mathcal{Q}}$, we obtain the one-channel effective Schr\"{o}dinger equation for the projected in state $\ket{\hat{\alpha}\Psi^{+}_{\alpha}}=\ket{\alpha}\braket{\alpha|\Psi^{+}_{\alpha}}$,
\tbox{Effective Schrödinger equation for in state}{
    (E_{\alpha}-\hat{H}_0 - \hat{\mathcal{W}}_{\alpha,\alpha}^{+})\ket{\hat{\alpha}\Psi^{+}_{\alpha}}=0\,, 
    \label{Sch_alpha}
}
with the complex potential
\begin{align}
\hat{\mathcal{W}}_{\alpha,\alpha}^{+} = \hat{V}_{\alpha,\alpha} + \hat{V}_{\alpha,\mathcal{Q}} \frac{1}{E_{\alpha}-\hat{H}_0-\hat{V}_{\mathcal{Q},\mathcal{Q}} + i\varepsilon } \hat{V}_{\mathcal{Q},\alpha}
\,.
\end{align}
On the other hand, the Feshbach projection of the two-channel Schr\"{o}dinger equation \eqref{couple_Sch} yields
\begin{align}
    (E_{\alpha}-\hat{H}_0-\hat{W}^{+}_{\alpha,\alpha})\ket{\hat{\alpha}\Psi^{+}_{\alpha}} - \hat{W}^{+}_{\alpha,\beta}\ket{\hat{\beta}\Psi^{+}_{\alpha}} &=0
    \,, \\
    -\hat{W}^{+}_{\beta,\alpha}\ket{\hat{\alpha} \Psi^{+}_{\alpha}}+(E_{\alpha}-\hat{H}_0-\hat{W}^{+}_{\beta,\beta}) \ket{\hat{\beta}\Psi^{+}_{\alpha}}&=0
    \,,
\end{align}
and then
\begin{align}
    \ket{\hat{\beta}\Psi^{+}_{\alpha}}=\frac{1}{E_{\alpha}-\hat{H}_0-\hat{W}^{\pm}_{\beta,\beta} +i\varepsilon}\hat{W}^{+}_{\beta,\alpha}\ket{\hat{\alpha} \Psi^{+}_{\alpha}}
    \,.
\end{align}
The $T$-matrix becomes
\begin{align}
T_{\beta,\alpha}
&=\braket{\beta|\hat{W}^+_{P,P}|\hat{P}\Psi^+_{\alpha}} 
\nn 
&=\braket{\beta|\hat{W}^+_{\beta,\alpha}|\hat{\alpha}\Psi^+_{\alpha}}  +  \braket{\beta|\hat{W}^+_{\beta,\beta}|\hat{\beta}\Psi^+_{\alpha}} 
\nn
&=\bra{\beta} \hat{W}^+_{\beta,\alpha}  + \hat{W}^+_{\beta,\beta} \frac{1}{E_{\alpha}-\hat{H}_0 - W^+_{\beta,\beta}+i\varepsilon} \hat{W}^+_{\beta,\alpha} \ket{\hat{\alpha}\Psi^+_{\alpha} }
\,.
\end{align}
To further rewrite this expression, we introduce the state $\bra{\psi^-_{\beta}\hat{\beta}}=\ket{\hat{\beta}\psi^-_{\beta}}^{\dagger}$ as the solution to the Lippmann-Schwinger equation
\begin{align}
\ket{\hat{\beta}\psi^{-}_{\beta}} &= \ket{\beta} + \frac{1}{E_{\alpha}-\hat{H}_0-\hat{W}^{-}_{\beta,\beta}- i\varepsilon} \hat{W}^{-}_{\beta,\beta} \ket{\beta}
\nn 
&=\ket{\beta} + \frac{1}{E_{\alpha}-\hat{H}_0- i\varepsilon} \hat{W}^{-}_{\beta,\beta} \ket{\hat{\beta}\psi^{-}_{\beta}}
\,. \label{beta_LS}
\end{align}
According to $-i\varepsilon$, the out state $\ket{\hat{\beta}\psi^-_{\beta}}$ must solve another one-channel Schr\"{o}dinger equation with the conjugate potential
\tbox{Effective Schrödinger equation for out state}{
    (E_{\alpha}-\hat{H}_0-\hat{W}_{\beta,\beta}^{-})\ket{\hat{\beta}\psi^{-}_{\beta}}=0
    \,,
    \label{Sch_beta}
}
under the outgoing boundary condition.
As a result, we obtain the following form of the $T$-matrix
\tbox{Twice-projected Feshbach formalism}{
T_{\beta,\alpha}
&= \braket{\psi^-_{\beta}\hat{\beta}|\hat{W}^+_{\beta,\alpha}|\hat{\alpha}\Psi^+_{\alpha} }
\,.
\label{Tbeta_alpha}
}
This equation looks close to the leading-order DWBA \eqref{LDWB}: the $T$-matrix is computed by sandwiching the in/out interacting states, which are the solutions to one-channel Schr\"{o}dinger equations \eqref{Sch_alpha} and \eqref{Sch_beta} under in/out boundary conditions. However, we stress that \eqref{Tbeta_alpha} is exact, and any perturbation theory has not been used so far.

In the Feshbach projection approach, we develop the perturbation theory of the effective potentials and solve the effective one-channel Schr\"{o}dinger equations non-perturbatively under truncated potentials. We recall that the necessary effective potentials to compute the $T$-matrix \eqref{Tbeta_alpha} are given by
\tbox{Effective potentials via Born series}{
\hat{\mathcal{W}}_{\alpha,\alpha}^+ &= \hat{V}_{\alpha,\alpha} + \hat{V}_{\alpha,\mathcal{Q}} \frac{1}{E_{\alpha}-\hat{H}_0-\hat{V}_{\mathcal{Q},\mathcal{Q}} + i\varepsilon } \hat{V}_{\mathcal{Q},\alpha}
=\hat{V}_{\alpha,\alpha} + \sum_{n=0}^{\infty}\hat{V}_{\alpha,\mathcal{Q}}\hat{G}_0^+(\hat{V}_{\mathcal{Q},\mathcal{Q}}\hat{G}_0^+)^n\hat{V}_{\mathcal{Q},\alpha}
    \,, \label{Wcal_alpha_alpha} \\
    \hat{W}^{+}_{\beta, \alpha} &=\hat{V}_{\beta,\alpha}+ \hat{V}_{\beta,Q}\frac{1}{E_{\alpha}-\hat{H}_0-\hat{V}_{Q,Q}+ i\varepsilon}\hat{V}_{Q,\alpha}
    =\hat{V}_{\beta,\alpha} + \sum_{n=0}^{\infty}\hat{V}_{\beta,Q}\hat{G}_0^+(\hat{V}_{Q,Q}\hat{G}_0^+)^n \hat{V}_{Q,\alpha}
    \,, \label{W_beta_alpha}
    \\
    \hat{W}^{+}_{\beta,\beta} &=\hat{V}_{\beta,\beta}+ \hat{V}_{\beta,Q}\frac{1}{E_{\alpha}-\hat{H}_0-\hat{V}_{Q,Q}+ i\varepsilon}\hat{V}_{Q,\beta}
    =\hat{V}_{\beta,\beta} + \sum_{n=0}^{\infty}\hat{V}_{\beta,Q}\hat{G}_0^+(\hat{V}_{Q,Q}\hat{G}_0^+)^n\hat{V}_{Q,\beta}
    \,,\label{W_beta_beta} 
}
with $\hat{Q}=\hat{1}-\hat{\alpha}-\hat{\beta}$ and $\hat{\mathcal{Q}}=\hat{\beta}+\hat{Q}=\hat{1}-\hat{\alpha}$. The minus potentials are their Hermitian conjugates. Although we can perturbatively compute the effective potentials via the Born series based on \eqref{Wcal_alpha_alpha}-\eqref{W_beta_beta}, in practice, the direct employment of \eqref{Wcal_alpha_alpha}-\eqref{W_beta_beta} will not be efficient. In Sec.~\ref{sec:resum}, we will elaborate on an alternative procedure by following the modern amplitude approach. In the rest of this section, we continue the formal discussion based on a given potential $\hat{V}$ to explain the underlying idea of our resummation.

\subsection{What did we sum up?}
\label{sec:example}
To have a better understanding of the formalisms introduced above, we consider a simple model and write the $T$-matrix in a diagrammatic way. We choose $\alpha$ as a two-massive-particle state while $\beta$ is a three-particle state with two-massive and one-massless particles. Our Hamiltonian is assumed to have two interactions 
\begin{align}
    V_{\bm{2',2}}=\braket{\bm{2'}|\hat{V}|\bm{2}}\,, \qquad V_{\bm{3,2}}=\braket{\bm{3}|\hat{V}|\bm{2}}
    \,.
    \label{V22_V32}
\end{align}
In this subsection, the bold numbers denote the number of particles, and the primes are added to distinguish different kinematics (non-bold numbers are left as particle labels in the next sections).
The first one $V_{\bm{2',2}}$ represents the potential force between two massive particles, whereas $V_{\bm{3,2}}$ describes a radiative process. To draw an analogy with the non-relativistic QED described by \eqref{QED_strong} and \eqref{QED_weak}, the former corresponds to the Coulomb potential $\Phi_{\rm Coulomb}(|\bm{x}_1-\bm{x}_2|)$, and the latter to $\frac{e_a}{m_a}\hat{\bm{A}}(\bm{x}_a)\cdot \hat{\bm{p}}_a$ (although in this case, the actual radiative process is $\bm{1}\to\bm{2}$, which appears as a disconnected process in $\bm{2}\to\bm{3}$).

Let us first revisit the Born series \eqref{Born_series}. Using the notation like \eqref{V22_V32}, the Born series for the $T$-matrix $T_{\bm{3,2}}=\braket{\bm{3}|\hat{T}|\bm{2}}$ is written as
\begin{align}
    T_{\bm{3,2}}=V_{\bm{3,2}}+\int\dd \Phi_{\bm{3'}} \frac{V_{\bm{3,3'}}V_{\bm{3',2}}}{E_{\bm 2}-E_{\bm{3'}}+i\varepsilon} + \int\dd \Phi_{\bm{2'}} \frac{V_{\bm{3,2'}}V_{\bm{2',2}}}{E_{\bm{2}}-E_{\bm{2'}}+i\varepsilon} + \cdots
    \,,
    \label{TB32}
\end{align}
where $\dd \Phi_{\bm n}$ is the phase-space integral of the $n$-particle state. Note that the 3-to-3 potential is non-zero due to the disconnected process
\begin{align}
    V_{\bm{3',3}}=2\omega \hdelta^{(3)}(\bm{k}'-\bm{k})V_{\bm{2',2}}
    \,,
\end{align}
where $\bm{k}$ and $\omega=|\bm{k}|$ are the three momentum and the frequency of the massless particle. Hence, the second term of \eqref{TB32} does not vanish even if the Hamiltonian does not have a contact 3-to-3 interaction.
We diagrammatically represent them as follows. The potentials are denoted by
\begin{align}
    V_{\bm{2',2}}=\Atwo{V}
    \,, \qquad
    V_{\bm{3,2}}=\Arad{V}
    \,, \qquad
    V_{\bm{3',3}}=
\begin{tikzpicture}[baseline=-2]
\begin{feynhand}
\threeL{-0.6}{0.6}{0.3}
\bub{0}{-0.3/2}{0.3}{V}{}
\end{feynhand}
\end{tikzpicture}
\,,
\end{align}
with the solid and wavy lines being massive and massless particles, respectively. The Born series \eqref{TB32} is then illustrated as
\begin{align}
    T_{\bm{3,2}}= \Arad{V}
    +
\begin{tikzpicture}[baseline=-2]
\begin{feynhand}
\def\y{0.3}
\def\segmentLength{1}
\pgfmathsetmacro{\xs}{0-0.8*\segmentLength}
\pgfmathsetmacro{\xA}{0}
\pgfmathsetmacro{\xB}{0+\segmentLength}
\pgfmathsetmacro{\xf}{0+\segmentLength+0.65*\segmentLength}
\threeL{\xs}{\xA}{\y}
\twoL{\xA}{\xf}{\y}
\bub{\xA}{0}{1.5*\y}{V}{}
\bub{\xB}{0}{\y}{V}{}
\end{feynhand}
\end{tikzpicture}
+
\begin{tikzpicture}[baseline=-2]
\begin{feynhand}
\def\y{0.3}
\def\segmentLength{1}
\pgfmathsetmacro{\xs}{0-0.65*\segmentLength}
\pgfmathsetmacro{\xA}{0}
\pgfmathsetmacro{\xB}{0+\segmentLength}
\pgfmathsetmacro{\xf}{0+\segmentLength+0.8*\segmentLength}
\threeL{\xs}{\xB}{\y}
\twoL{\xB}{\xf}{\y}
\bub{\xA}{-\y/2}{\y}{V}{}
\bub{\xB}{0}{1.5*\y}{V}{}
\end{feynhand}
\end{tikzpicture}
+\cdots
\,.
\end{align}
The products of diagrams are understood as the phase-space integrals multiplied by the energy denominator, as shown in \eqref{TB32}. We do not consider diagrams with more than two massive lines because we are interested in the classical regime where pair creation/annihilation of massive particles would be suppressed. In other words, while the diagrams $V_{\bm{2,2}}$ and $V_{\bm{3,2}}$ are taken into account, their crossed diagrams like $V_{\bm{4,1}}$ (four-massive and one-massless) are neglected.

We use this diagrammatic notation to understand what types of diagrams are summed up within the DWBA and the Feshbach projection. For the former, we choose $V_{\bm{2',2}}$ as the strong interaction and $V_{\bm{3,2}}$ as the weak interaction, respectively. The leading-order DWBA is given in \eqref{LDWB}. Expanding it using the Born series, we see that the leading DWBA sums up the following diagrams
\begin{align}
    \overset{(1)}{T}{}^{{\rm DWBA}}_{\bm{3,2}}
    &=
    \sum_{i,j}
\begin{tikzpicture}[baseline=-2]
\begin{feynhand}
\def\y{0.3}
\def\segmentLength{0.8}
\pgfmathsetmacro{\xs}{0-0.5*\segmentLength}
\pgfmathsetmacro{\xA}{0+0.15}
\pgfmathsetmacro{\xB}{0+\segmentLength}
\pgfmathsetmacro{\xC}{0+2*\segmentLength-0.15}
\pgfmathsetmacro{\xD}{0+3*\segmentLength}
\pgfmathsetmacro{\xE}{0+4*\segmentLength+0.15}
\pgfmathsetmacro{\xF}{0+5*\segmentLength}
\pgfmathsetmacro{\xG}{0+6*\segmentLength-0.15}
\pgfmathsetmacro{\xf}{0+6*\segmentLength+0.5*\segmentLength}
\threeL{\xs}{\xD}{\y}
\twoL{\xD}{\xf}{\y}
\bub{\xA}{-\y/2}{\y}{V}{}
\bub{\xC}{-\y/2}{\y}{V}{}
\bub{\xD}{0}{1.5*\y}{V}{}
\bub{\xE}{0}{\y}{V}{}
\bub{\xG}{0}{\y}{V}{}
\node at (\xB,-\y/2) {$\cdots$};
\node at (\xF,0) {$\cdots$};
\draw[decorate,decoration={brace,mirror,amplitude=0.1cm}] 
    (\xs, -\y-0.2) -- (\xD-1.5*\y, -\y-0.2);
\node at ({(\xD-1.5*\y-\xs)/2+\xs}, -\y-0.5) {$i$};
\draw[decorate,decoration={brace,mirror,amplitude=0.1cm}] 
    (\xD+1.5*\y, -\y-0.2) -- (\xf, -\y-0.2);
\node at ({(\xf-\xD-1.5*\y)/2+\xD+1.5*\y}, -\y-0.5) {$j$};
\end{feynhand}
\end{tikzpicture}
\nn
&=\sum_{i,j}
\begin{tikzpicture}[baseline=-2]
\begin{feynhand}
\def\y{0.3}
\def\segmentLength{1}
\pgfmathsetmacro{\xs}{0-0.5*\segmentLength}
\pgfmathsetmacro{\xA}{0}
\pgfmathsetmacro{\xB}{0+\segmentLength}
\pgfmathsetmacro{\xC}{0+2*\segmentLength}
\pgfmathsetmacro{\xf}{0+2*\segmentLength+0.5*\segmentLength}
\threeL{\xs}{\xB}{\y}
\twoL{\xB}{\xf}{\y}
\bub{\xA}{-\y/2}{\y}{V^i}{}
\bub{\xB}{0}{1.5*\y}{V}{}
\bub{\xC}{0}{\y}{V^j}{}
\end{feynhand}
\end{tikzpicture}
\,.
\label{LDWBA}
\end{align}
The symbol $V^i$ denotes the $i$-th product of the same diagrams. The quadratic order in $V_{\bm{3,2}}$ is absent. Hence, the next-to-leading order DWBA is
\begin{align}
     \overset{(3)}{T}{}^{{\rm DWBA}}_{\bm{3,2}} 
     &=
    \sum_{i,j,k,l}
\begin{tikzpicture}[baseline=-2]
\begin{feynhand}
\def\y{0.3}
\def\segmentLength{1}
\pgfmathsetmacro{\xs}{0-0.5*\segmentLength}
\pgfmathsetmacro{\xA}{0}
\pgfmathsetmacro{\xB}{0+\segmentLength}
\pgfmathsetmacro{\xC}{0+2*\segmentLength}
\pgfmathsetmacro{\xD}{0+3*\segmentLength}
\pgfmathsetmacro{\xE}{0+4*\segmentLength}
\pgfmathsetmacro{\xF}{0+5*\segmentLength}
\pgfmathsetmacro{\xG}{0+6*\segmentLength}
\pgfmathsetmacro{\xf}{0+6*\segmentLength+0.5*\segmentLength}
\threeL{\xs}{\xB}{\y}
\twoL{\xB}{\xD}{\y}
\threeL{\xD}{\xF}{\y}
\twoL{\xF}{\xf}{\y}
\bub{\xA}{-\y/2}{\y}{V^i}{}
\bub{\xB}{0}{1.5*\y}{V}{}
\bub{\xC}{0}{\y}{V^j}{}
\bub{\xD}{0}{1.5*\y}{V}{}
\bub{\xE}{-\y/2}{\y}{V^k}{}
\bub{\xF}{0}{1.5*\y}{V}{}
\bub{\xG}{0}{\y}{V^l}{}
\end{feynhand}
\end{tikzpicture}
\nn
&+
    \sum_{i,j,k,l}
\begin{tikzpicture}[baseline=-2]
\begin{feynhand}
\def\y{0.3}
\def\segmentLength{1}
\pgfmathsetmacro{\xs}{0-0.5*\segmentLength}
\pgfmathsetmacro{\xA}{0}
\pgfmathsetmacro{\xB}{0+\segmentLength}
\pgfmathsetmacro{\xC}{0+2*\segmentLength}
\pgfmathsetmacro{\xD}{0+3*\segmentLength}
\pgfmathsetmacro{\xE}{0+4*\segmentLength}
\pgfmathsetmacro{\xF}{0+5*\segmentLength}
\pgfmathsetmacro{\xG}{0+6*\segmentLength}
\pgfmathsetmacro{\xf}{0+6*\segmentLength+0.5*\segmentLength}
\propag[boson] (\xs,\y) -- (\xF, \y);
\twoLlow{\xs}{\xB}{\y}
\threeLlow{\xB}{\xD}{\y}
\twoLlow{\xD}{\xF}{\y}
\twoL{\xF}{\xf}{\y}
\bub{\xA}{-\y/2}{\y}{V^i}{}
\bub{\xB}{-\y/2}{1.2*\y}{V}{}
\bub{\xC}{-2*\y/3}{0.8*\y}{V^j\,}{\scriptsize}
\bub{\xD}{-\y/2}{1.2*\y}{V}{}
\bub{\xE}{-\y/2}{\y}{V^k}{}
\bub{\xF}{0}{1.5*\y}{V}{}
\bub{\xG}{0}{\y}{V^l}{}
\end{feynhand}
\end{tikzpicture}
\nn
&+
    \sum_{i,j,k,l}
\begin{tikzpicture}[baseline=-2]
\begin{feynhand}
\def\y{0.3}
\def\segmentLength{1}
\pgfmathsetmacro{\xs}{0-0.5*\segmentLength}
\pgfmathsetmacro{\xA}{0}
\pgfmathsetmacro{\xB}{0+\segmentLength}
\pgfmathsetmacro{\xC}{0+2*\segmentLength}
\pgfmathsetmacro{\xD}{0+3*\segmentLength}
\pgfmathsetmacro{\xE}{0+4*\segmentLength}
\pgfmathsetmacro{\xF}{0+5*\segmentLength}
\pgfmathsetmacro{\xG}{0+6*\segmentLength}
\pgfmathsetmacro{\xf}{0+6*\segmentLength+0.5*\segmentLength}
\propag[boson] (\xs,\y+0.2) -- (\xD-0.1, 0.2);
\propag[boson] (\xB+0.1, 0.2) -- (\xF-0.1, \y+0.1);
\twoLlow{\xs}{\xB}{\y}
\draw (\xB,-\y) -- (\xD,-\y);
\draw (\xB,-\y/3) -- (\xD,-\y/3);
\twoLlow{\xD}{\xF}{\y}
\twoL{\xF}{\xf}{\y}
\bub{\xA}{-\y/2}{\y}{V^i}{}
\bub{\xB}{-\y/2}{1.2*\y}{V}{}
\bub{\xC}{-2*\y/3}{0.8*\y}{V^j\,}{\scriptsize}
\bub{\xD}{-\y/2}{1.2*\y}{V}{}
\bub{\xE}{-\y/2}{\y}{V^k}{}
\bub{\xF}{0}{1.5*\y}{V}{}
\bub{\xG}{0}{\y}{V^l}{}
\end{feynhand}
\end{tikzpicture}
\,.
\label{NLO_DWBA}
\end{align}
They are all orders in $V_{\bm{2',2}}$ and perturbative only in $V_{\bm{3,2}}$.

The Feshbach projection formalism resums the diagrams differently. The perturbative expansion of the effective potentials \eqref{Wcal_alpha_alpha}-\eqref{W_beta_beta} is diagrammatically given by
\begin{align}
    \Atwo{\mathcal{W}} = \Atwo{V} + 
\begin{tikzpicture}[baseline=-2]
\begin{feynhand}
\def\y{0.3}
\def\segmentLength{1.2}
\pgfmathsetmacro{\xs}{0-0.5*\segmentLength-0.1}
\pgfmathsetmacro{\xA}{0}
\pgfmathsetmacro{\xB}{0+\segmentLength}
\pgfmathsetmacro{\xf}{0+\segmentLength+0.5*\segmentLength+0.1}
\twoL{\xs}{\xA}{\y}
\threeL{\xA}{\xB}{\y}
\twoL{\xB}{\xf}{\y}
\bub{\xA}{0}{1.5*\y}{V}{}
\bub{\xB}{0}{1.5*\y}{V}{}
\end{feynhand}
\end{tikzpicture}
+
\begin{tikzpicture}[baseline=-2]
\begin{feynhand}
\def\y{0.3}
\def\segmentLength{1}
\pgfmathsetmacro{\xs}{0-0.5*\segmentLength-0.2}
\pgfmathsetmacro{\xA}{0}
\pgfmathsetmacro{\xB}{0+\segmentLength}
\pgfmathsetmacro{\xC}{0+2*\segmentLength}
\pgfmathsetmacro{\xf}{0+2*\segmentLength+0.5*\segmentLength+0.2}
\twoL{\xs}{\xA}{\y}
\threeL{\xA}{\xC}{\y}
\twoL{\xC}{\xf}{\y}
\bub{\xA}{0}{1.5*\y}{V}{}
\bub{\xB}{-\y/2}{\y}{V}{}
\bub{\xC}{0}{1.5*\y}{V}{}
\end{feynhand}
\end{tikzpicture}
+
\cdots
\,,
\label{Wcal_diagram}
\end{align}
and
\begin{align}
    \Arad{W}&=\Arad{V}
    \,, 
    \\
    \begin{tikzpicture}[baseline=-2]
\begin{feynhand}
\threeL{-0.8}{0.8}{0.3}
\bub{0}{0}{1.5*0.3}{W}{}
\end{feynhand}
\end{tikzpicture}
&=
\begin{tikzpicture}[baseline=-2]
\begin{feynhand}
\threeL{-0.6}{0.6}{0.3}
\bub{0}{-0.3/2}{0.3}{V}{}
\end{feynhand}
\end{tikzpicture}
+
\begin{tikzpicture}[baseline=-2]
\begin{feynhand}
\def\y{0.3}
\def\segmentLength{1.2}
\pgfmathsetmacro{\xs}{0-0.5*\segmentLength}
\pgfmathsetmacro{\xA}{0}
\pgfmathsetmacro{\xB}{0+\segmentLength}
\pgfmathsetmacro{\xf}{0+\segmentLength+0.5*\segmentLength}
\propag[boson] (\xs, \y) -- (\xf, \y);
\twoLlow{\xs}{\xA}{\y}
\threeLlow{\xA}{\xB}{\y}
\twoLlow{\xB}{\xf}{\y}
\bub{\xA}{-\y/2}{1.2*\y}{V}{}
\bub{\xB}{-\y/2}{1.2*\y}{V}{}
\end{feynhand}
\end{tikzpicture}
+
\begin{tikzpicture}[baseline=-2]
\begin{feynhand}
\def\y{0.3}
\def\segmentLength{0.8}
\pgfmathsetmacro{\xs}{0-0.5*\segmentLength-0.2}
\pgfmathsetmacro{\xA}{0}
\pgfmathsetmacro{\xB}{0+\segmentLength}
\pgfmathsetmacro{\xC}{0+2*\segmentLength}
\pgfmathsetmacro{\xf}{0+2*\segmentLength+0.5*\segmentLength+0.2}
\propag[boson] (\xs, \y) -- (\xf, \y);
\twoLlow{\xs}{\xA}{\y}
\threeLlow{\xA}{\xC}{\y}
\twoLlow{\xC}{\xf}{\y}
\bub{\xA}{-\y/2}{1.2*\y}{V}{}
\bub{\xB}{-2*\y/3}{0.8*\y}{V~}{\scriptsize}
\bub{\xC}{-\y/2}{1.2*\y}{V}{}
\end{feynhand}
\end{tikzpicture}
+
\cdots
\nn
&+
\begin{tikzpicture}[baseline=-2]
\begin{feynhand}
\def\y{0.3}
\def\segmentLength{1.2}
\pgfmathsetmacro{\xs}{0-0.5*\segmentLength}
\pgfmathsetmacro{\xA}{0}
\pgfmathsetmacro{\xB}{0+\segmentLength}
\pgfmathsetmacro{\xf}{0+\segmentLength+0.5*\segmentLength}
\propag[boson] (\xs, \y+0.1) -- (\xB, 0+0.15);
\propag[boson] (\xf, \y+0.1) -- (\xA, 0+0.15);
\twoLlow{\xs}{\xA}{\y}
\draw (\xA,-\y) -- (\xB,-\y);
\draw (\xA,-\y/3) -- (\xB,-\y/3);
\twoLlow{\xB}{\xf}{\y}
\bub{\xA}{-\y/2}{1.2*\y}{V}{}
\bub{\xB}{-\y/2}{1.2*\y}{V}{}
\end{feynhand}
\end{tikzpicture}
+\cdots
\nn
&=
\begin{tikzpicture}[baseline=-2]
\begin{feynhand}
\threeL{-0.6}{0.6}{0.3}
\bub{0}{-0.3/2}{0.3}{\mathcal{W}}{}
\end{feynhand}
\end{tikzpicture}
+
\begin{tikzpicture}[baseline=-2]
\begin{feynhand}
\threeL{-0.8}{0.8}{0.3}
\bub{0}{0}{1.5*0.3}{W_c}{}
\end{feynhand}
\end{tikzpicture}
\,,
\label{W3_diagram}
\end{align}
where $W_c$ denotes the connected part, like the middle line of \eqref{W3_diagram}.
Recall $\hat{\mathcal{Q}}=\hat{1}-\ket{\bm{2}}\bra{\bm{2}}$ and $\hat{Q}=\hat{1}-\ket{\bm{2}}\bra{\bm{2}}-\ket{\bm{3}}\bra{\bm{3}}$. Therefore, the internal lines of $\mathcal{W}$ have to be at least three, and those of $W$ are at least four. In this example, $W^+_{\bm{3,2}}$ agrees with $V_{\bm{3,2}}$ due to the absence of $V_{\bm{n,2}}$ with $n\geq 4$, but this is an artefact of the simplified setup. Note also that $W_{\bm{3',3}}^+$ with a disconnected radiation has the same structure as $\mathcal{W}^+_{\bm{2',2}}$; after the radiation is factored out, only the two-massive-particle state is excluded from the internal lines connecting the blobs, which is precisely the structure of $\mathcal{W}^+_{\bm{2}',\bm{2}}$. This is a generic feature of the 3-to-3 potential, not due to the simplification here.

The connected diagrams $W_c$, or generically speaking, diagrams interacting with the massless state in $W_{\bm{3}',\bm{3}}$ can be understood as yielding a deviation from the free propagation of the radiation. In the present paper, we only focus on the generation of radiation and do not consider these contributions explicitly.

Taking the free propagation approximation, the $T$-matrix in the Feshbach formalism is written as
\begin{align}
T_{\bm{3,2}}^{\rm F}=
\sum_{i,j}
\begin{tikzpicture}[baseline=-2]
\begin{feynhand}
\def\y{0.3}
\def\segmentLength{1}
\pgfmathsetmacro{\xs}{0-0.5*\segmentLength}
\pgfmathsetmacro{\xA}{0}
\pgfmathsetmacro{\xB}{0+\segmentLength}
\pgfmathsetmacro{\xC}{0+2*\segmentLength}
\pgfmathsetmacro{\xf}{0+2*\segmentLength+0.5*\segmentLength}
\threeL{\xs}{\xB}{\y}
\twoL{\xB}{\xf}{\y}
\bub{\xA}{-\y/2}{\y}{\mathcal{W}^i~}{\small}
\bub{\xB}{0}{1.5*\y}{W}{}
\bub{\xC}{0}{\y}{\mathcal{W}^j~}{\small}
\end{feynhand}
\end{tikzpicture}
\,.
\end{align}
It is now transparent that this expression of the $T$-matrix sums up the diagrams with repeated topologies. At the leading order, the 2-to-2 effective potential agrees with the original 2-to-2 potential, so the result agrees with the leading DWBA \eqref{LDWBA} where the ladder topologies before and after the radiation emission are resummed. They differ at the next-to-leading order. The DWBA \eqref{NLO_DWBA} adds up the ladders of $V_{\bm{2',2}}$ at each stage of the transition processes $\bm{2}\leftrightarrow \bm{3}$, but the transition processes $\bm{2}\leftrightarrow \bm{3}$ occur only a finite number of times. On the other hand, the Feshbach formalism replaces the ladder with one including internal exchanges of massless states, the second term onwards in \eqref{Wcal_diagram}. Then, this effective ladder is summed up an infinite number of times. Hence, the perturbation theory based on the Feshbach projection can be regarded as a perturbation theory based on the topology of diagrams: at each order of perturbation, the number of types of internal topology is finite, but diagrams made by repeating such internal topologies are all added.

It would depend on the problems for which of the DWBA or the Feshbach formalism gives a good approximation to describe non-perturbative effects. The DWBA should be powerful if the ``strong'' and ``weak'' interactions can be clearly separated and the ``strong'' Schr\"{o}dinger equation can be easily solved. The typical example is the non-relativistic QED, where the solution is given by the Coulomb wavefunction. On the other hand, in the Feshbach formalism, both Coulomb and Lorentz forces can be treated equally as they are packaged into the effective potential. It does not require splitting the ``potential'' and ``radiation'' modes (because any notion of the field is not necessary) and can be applied even in relativistic systems. In fact, we will see shortly that, applying it to gravity, the 2-to-2 effective potential $\mathcal{W}_{\bm{2}',\bm{2}}$ corresponds to the post-Minkowskian two-body potential valid to all orders in velocities~\cite{Cheung:2018wkq}. The perturbative expansion of the potentials should thus be the $1/r$ expansion in the position space. This would be an appropriate perturbation scheme for classical dynamics as long as the distance between the two objects remains sufficiently large during their time evolution.


\section{Resumming amplitudes in classical regime}
\label{sec:resum}
In the previous section, we introduced a framework based on the Feshbach projection suitable for computing non-perturbative amplitudes in the post-Minkowskian regime. It requires obtaining three effective potentials, originally defined in \eqref{Wcal_alpha_alpha}-\eqref{W_beta_beta}, solving the effective Schr\"{o}dinger equations \eqref{Sch_alpha} and \eqref{Sch_beta}, and then computing the $T$-matrix \eqref{Tbeta_alpha}. They are still tremendous tasks, and this section explains how we can carry them out.

In the following, we restrict our attention to two-body gravitational scattering with one-graviton emission ($\alpha=$ two-massive-particle state and $\beta=$ two-massive-particle and one-graviton state), although the framework can be applied to generic inelastic processes. The particles are labelled by
\begin{align}
\begin{tikzpicture}[baseline=-2]
\begin{feynhand}
\def\y{0.5}
\def\segmentLength{1.5}
\pgfmathsetmacro{\xs}{0-0.8*\segmentLength}
\pgfmathsetmacro{\xA}{0}
\pgfmathsetmacro{\xf}{0+0.8*\segmentLength}
\draw (\xs, 0) node[left] {$p_1'$}  -- (\xA, 0);
\draw (\xs, -\y)node[left] {$p_2'$}  -- (\xA, -\y);
\propag[boson] (\xs, \y) node[left] {$k$} -- (\xA, \y);
\draw (\xA, \y/2) -- (\xf, \y/2) node[right] {$p_1$};
\draw (\xA, -\y/2) -- (\xf, -\y/2) node[right] {$p_2$};
\bub{\xA}{0}{1.5*\y}{}{}
\end{feynhand}
\end{tikzpicture}
.
\end{align}
The graviton four-momentum is denoted by $k^{\mu}=(\omega,\bm{k})$, whereas the momenta of massive particles are $p_1^{\mu}=(E_1,\bm{p}_1),p_1'{}^{\mu}=(E_1',\bm{p}_1')$, and so on. We sometimes use $\bml$ and $\bmr$ for massless and massive momenta, especially when they are integrated. The distinction between massless and massive momenta is important when we recover $\hbar$; they must scale as $\bmp,\bmr =\mathcal{O}(\hbar^0)$ and $\bmk,\bml=\mathcal{O}(\hbar)$ in the classical limit $\hbar \to 0$.

Our strategy follows recent developments of the on-shell program. We shall take perturbative scattering amplitudes as our starting point, rather than the Hamiltonian or fields, by intending to apply various techniques developed for perturbative amplitude calculations. Once the amplitudes are given, one can map them to potentials by following EFT ideas~\cite{Neill:2013wsa, Cheung:2018wkq}, for example. The point is that we now view these potentials as the effective potentials of the Schr\"{o}dinger equation (or the Lippmann-Schwinger equation) in the sense of Feshbach, not the EFT. This idea was first proposed in \cite{Iwasaki:1971iy, Iwasaki:1971vb, Cristofoli:2019neg}, although its connection to the projector theory of Feshbach was not realised. This new perspective allows us to extend the mapping between the amplitudes and the potentials into radiative processes. The remaining tasks are then computing the Schr\"{o}dinger equations and the $T$-matrix. The classical limit simplifies these problems because solving the Schr\"{o}dinger equation in the classical limit can be essentially reduced to solving a classical equation of motion. Let us elaborate on them in order.

\subsection{From perturbative amplitudes to effective potentials}
In Sec.~\ref{sec:Feshbach}, we defined the effective potentials from the Hamiltonian of a ``fundamental'' theory. This section describes a way for providing these effective potentials without the use of a ``fundamental'' Hamiltonian.

Let us introduce the $T$-operator by the Born series
\begin{align}
    \hat{T}=\sum_{n=0}^{\infty}\hat{V}(\hat{G}_0^+\hat{V})^n = \hat{V}+\hat{V}\hat{G}_0^+\hat{V}+\cdots
    \,,
\end{align}
which computes any scattering processes by sandwiching free states, e.g.,
\begin{align}
    T_{\beta,\alpha}=\braket{\beta|\hat{T}|\alpha}
    \,.
\end{align}
On the other hand, what we need here is the projected $T$-operators
\begin{align}
    \hat{T}_{P,P}:=\hat{P}\hat{T}\hat{P}\,, \qquad \hat{T}_{\alpha,\alpha}:=\hat{\alpha}\hat{T}\hat{\alpha}
    \,,
\end{align}
where we recall $\hat{P}=\hat{\alpha}+\hat{\beta}$. By definition, the projected $T$-operators only compute the selected scattering processes and give zero otherwise. The projector theory of Feshbach claimed that the projected $T$-operators are given by the Born series of the effective potentials
\begin{align}
    \hat{T}_{P,P}&=\sum_{n=0}^{\infty}\hat{W}^+_{P,P}(\hat{G}_0^+ \hat{W}^+_{P,P})^n = \hat{W}_{P,P}+\hat{W}_{P,P}\hat{G}_0^+\hat{W}_{P,P}+\cdots
    \,, \\
    \hat{T}_{\alpha,\alpha}&=\sum_{n=0}^{\infty}\hat{\mathcal{W}}^+_{\alpha,\alpha}(\hat{G}_0^+ \hat{\mathcal{W}}^+_{\alpha,\alpha})^n = \hat{\mathcal{W}}_{\alpha,\alpha}+\hat{\mathcal{W}}_{\alpha,\alpha}\hat{G}_0^+\hat{\mathcal{W}}_{\alpha,\alpha}+\cdots
    \,.
\end{align}

We can perturbatively invert these Born series to obtain the inverse formula
\tbox{Born subtraction}{
\hat{V}&=\sum_{n=0}^{\infty}\hat{T}(-\hat{G}_0^+ \hat{T})^n=\hat{T}-\hat{T}\hat{G}_0^+\hat{T}+\cdots
\,, \label{Bsub}
\\
    \hat{W}^+_{P,P}&=\sum_{n=0}^{\infty}\hat{T}_{P,P}(-\hat{G}_0^+ \hat{T}_{P,P})^n=\hat{T}_{P,P}-\hat{T}_{P,P}\hat{G}_0^+\hat{T}_{P,P}+\cdots
    \,, 
    \label{Bsub_P}\\
    \hat{\mathcal{W}}^+_{\alpha,\alpha}&=\sum_{n=0}^{\infty}\hat{T}_{\alpha,\alpha}(-\hat{G}_0^+ \hat{T}_{\alpha,\alpha})^n
    =\hat{T}_{\alpha,\alpha}-\hat{T}_{\alpha,\alpha}\hat{G}_0^+\hat{T}_{\alpha,\alpha}+\cdots
    \,,
    \label{Bsub_alpha}
}
which we call Born subtractions by following~\cite{Cristofoli:2019neg}. Eqs.~\eqref{Bsub_P} and \eqref{Bsub_alpha} can be regarded as {\it defining} the effective potentials by the $T$-matrices, hence the scattering amplitudes, without reference to the fundamental Hamiltonian. The distinction between the fundamental and effective potentials is seen in whether all internal channels must be subtracted in non-linear terms or only selected channels are subtracted. In other words, the fundamental potential \eqref{Bsub} requires knowing all scattering processes, whereas the effective potentials \eqref{Bsub_P} and \eqref{Bsub_alpha} are computed by only selected scattering processes, $P\to P'$ and $\alpha \to \alpha'$.

In Ref.~\cite{Cristofoli:2019neg}, one of the authors showed that the one-channel potential \eqref{Bsub_alpha} agrees with the two-body gravitational potential in the EFT approach~\cite{Cheung:2018wkq}. The two-channel Hamiltonian with $\hat{W}^+_{P,P}$ can be interpreted as an extension of the EFT to incorporate the radiative state $\beta$. However, the 2-to-2 effective potential in the two-channel Hamiltonian,
\begin{align}
    \hat{W}_{\alpha,\alpha}^+=\hat{T}_{\alpha,\alpha} -\hat{T}_{\alpha,P}\hat{G}_0^+\hat{T}_{P,\alpha}+\cdots
    =\hat{T}_{\alpha,\alpha} -\hat{T}_{\alpha,\alpha}\hat{G}_0^+\hat{T}_{\alpha,\alpha}  -\hat{T}_{\alpha,\beta}\hat{G}_0^+\hat{T}_{\beta,\alpha}+\cdots
    \,,
\end{align}
is {\it different} from the one-channel potential $\hat{\mathcal{W}}_{\alpha,\alpha}^+$ due to the additional subtraction of the $\beta$-channel. The necessity of the additional subtraction is understood because this effective theory has the 2-to-3 potential, which eventually generates a force between two bodies by exchanging the graviton; hence, this contribution has to be subtracted in the two-body potential to avoid double-counting the same effect. The resultant effective theory of two channels would thus have a similar difficulty to the non-relativistic QED. On the other hand, the twice-projection of Feshbach \eqref{Tbeta_alpha} yields the $T$-matrix being computed by {\it both} one-channel potential $\hat{\mathcal{W}}^+_{\alpha,\alpha}$ and two-channel potentials $\hat{W}^+_{\beta,\alpha}$ and $\hat{W}^+_{\beta,\beta}$. The last potential can be reduced to the one-channel potential $\hat{\mathcal{W}}^+_{\alpha,\alpha}$ under the free propagation approximation, as we have seen in Sec.~\ref{sec:example}. Therefore, the twice-projection allows us to discuss radiative processes while reusing the computations of the gravitational potential in the post-Minkowskian (PM) expansion. Hereafter, we denote
\begin{align}
    \hat{V}_{\rm PM}:=\hat{\mathcal{W}}_{\alpha,\alpha}^+\,, \qquad \hat{R}_{\rm PM}:=\hat{W}_{\beta,\alpha}^+
    \,,
    \label{PM_potentials}
\end{align}
and call them the two-body potential and the radiation potential, respectively. The momentum-space representations of these potentials are denoted by
\tbox{Two-body and radiation potentials}{
    \braket{p_1';p_2'|\hat{V}_{\rm PM}|p_1;p_2} &= \sqrt{16E_1 E_2 E_1 ' E_2'}\,\hdelta^{(3)}(\bmp_1'+\bmp_2'-\bmp_1 - \bmp_2) V_{\rm PM} \,, 
    \label{PM_potential} \\
    \braket{p_1';p_2';k|\hat{R}_{\rm PM}|p_1;p_2} &= \sqrt{32E_1 E_2 E_1 ' E_2'\omega}\,\hdelta^{(3)}(\bmp_1'+\bmp_2'+\bmk-\bmp_1 - \bmp_2) R_{\rm PM}\,,
    \label{R_potential}
}
where the delta functions guarantee the momentum conservation, and the overall energy factors are added for later convenience.

Since the present paper focuses on the development of the formalism, we explicitly calculate the radiation potential only at the leading order to avoid technical complications due to higher-order effects. The $T$-matrix is split into the disconnected and connected parts,
\begin{align}
\Arad{T}  &= 
\begin{tikzpicture}[baseline=-2]
    \begin{feynhand}
        \twoL{-0.6}{0.6}{0.5}
        \propag[boson] (-0.6, 0.5)--(0, 0.25);
    \end{feynhand}
    \end{tikzpicture}
    +
    \begin{tikzpicture}[baseline=-2]
    \begin{feynhand}
        \twoL{-0.6}{0.6}{0.5}
        \propag[boson] (-0.6, -0.5)--(0, -0.25);
    \end{feynhand}
    \end{tikzpicture}
    +
    \Arad{T_c}
    \,,
\end{align}
which are respectively given by
\begin{align}
\begin{tikzpicture}[baseline=-2]
    \begin{feynhand}
        \twoL{-0.6}{0.6}{0.5}
        \propag[boson] (-0.6, 0.5)--(0, 0.25);
    \end{feynhand}
    \end{tikzpicture}
    +
    \begin{tikzpicture}[baseline=-2]
    \begin{feynhand}
        \twoL{-0.6}{0.6}{0.5}
        \propag[boson] (-0.6, -0.5)--(0, -0.25);
    \end{feynhand}
    \end{tikzpicture}
        &=\hdelta^{(3)}(\bm{p}'_1+\bm{p}'_2+\bm{k}-\bm{p}_1-\bm{p}_2) 
     \Big[ \kappa (\varepsilon \cdot p_1)^2 \braket{p_2'|p_2} +\kappa (\varepsilon \cdot p_2)^2 \braket{p_1'|p_1} \Big] ,
     \\
\Arad{T_c}
&=-\hdelta^{(3)}(\bm{p}'_1+\bm{p}'_2+\bm{k}-\bm{p}_1-\bm{p}_2) \Amp_{\bm{3},\bm{2}} = \mathcal{O}(\kappa^3)
    \,.
\end{align}
Here, $\varepsilon_{\mu}=\varepsilon_{\mu}(k)$ is the polarisation vector with the polarisation being implicit, and $\Amp_{\bm{3},\bm{2}}$ is an off-energy-shell extension of the 2-to-3 amplitude.\footnote{The overall minus sign is due to the different sign conventions of the $T$-matrix and the amplitudes: $S_{\beta,\alpha}=\braket{\beta|\alpha}-i\hdelta(E_{\alpha}-E_{\beta})T_{\beta,\alpha}=\braket{\beta|\alpha}+i\hdelta^{(4)}(p_{\alpha}-p_{\beta})\Amp_{\beta,\alpha}$} We stress that the $T$-matrix is defined after removing the energy conservation (see \eqref{S-matrix}), and only has the delta functions for the momentum conservation. The three-point does not vanish even for the real kinematics. Hence, the $T$-matrix solely comes from the disconnected part at the leading order in $\kappa=\sqrt{32\pi G}$. The radiation potential agrees with the $T$-matrix at the leading order, yielding
\tbox{Leading-order radiation potential in momentum space}{
R_{\rm PM}= \frac{\kappa}{\sqrt{2\omega}} \left[ \hdelta^{(3)}(\bmp_2'-\bmp_2) \frac{(\varepsilon \cdot p_1')^2}{\sqrt{4E_1 E_1'}} + \hdelta^{(3)}(\bmp_1'-\bmp_1) \frac{(\varepsilon \cdot p_2')^2}{\sqrt{4E_2 E_2'}}\right]  +\mathcal{O}(\kappa^3)
\,.
\label{Rmomentum_LO}
}

\subsection{Effective potentials to wavefunctions}
Having obtained the effective potentials through the Born subtractions, the next problem is solving the Schr\"{o}dinger equations \eqref{Sch_alpha} and \eqref{Sch_beta}. Since we are interested in the classical regime, the powerful method to compute a non-perturbative wavefunction is the WKB approximation.

We factorise the two-particle states into the centre-of-mass (COM) motion and relative motion states
\tbox{Two-particle states}{
\ket{p_1;p_2}=\sqrt{4E_1E_2}\ket{\bm{P}}\otimes\ket{\bm{p}}\,,\qquad
    \ket{\hat{\alpha} \Psi^{\pm}_{12}}=\sqrt{4E_1E_2}\ket{\bm{P}}\otimes \ket{\Psi^{\pm}_{\bm p}}
    \label{two-body_state}
}
with the non-relativistic normalisations for ``bold'' states
\begin{align}
    \braket{\bmP'|\bmP}=\hdelta^{(3)}(\bm{P}'-\bm{P})\,, \qquad
    \braket{\bmp'|\bmp}=\hdelta^{(3)}(\bm{p}'-\bm{p})\,.
    \label{NR_normal}
\end{align}
The total and relative momenta are defined by
\begin{align}
    \bm{P}=\bm{p}_1+\bm{p}_2\,, \qquad \bm{p}=\eta_2 \bm{p}_1-\eta_1 \bm{p}_2
    \,,
    \label{defPp}
\end{align}
with $\eta_1(\bmp_1,\bmp_2)+\eta_2(\bmp_1,\bmp_2)=1$. Note that there is no unique notion of relativistic COM. This ambiguity usually does not enter into the computation of the 2-to-2 amplitude by working in the COM frame $\bmP=\bmp_1+\bmp_2=0$, where $\bmp=\bmp_1=-\bmp_2$ irrespectively of the choice of $\eta_a$. In the present paper, however, we eventually require a deviation from the COM frame due to the presence of graviton. We leave $\eta_a$ undecided for a while, and will choose it for our convenience later. 

In our case, we should also deal with the three-particle states. While the COM motion of the three particles can be factored out, the remaining relative motion is still a two-body state, which cannot be easily solved when interacting. However, the gravitational force between the graviton and massive bodies should be much weaker than the force between the massive bodies. The relative motion state may be approximately factorised into the relative motion of the massive bodies and the motion of the graviton, a la the Born-Oppenheimer approximation. Therefore, our three-particle state may be approximated by
\begin{tcolorbox}[top=-5pt,colback=black!5!white,colframe=black!75!white,title=
Three-particle states]
\begin{alignat}{4}
    \ket{p_1;p_2;k}&=\sqrt{8E_1E_2 \omega}
    &&
    ~~~\ket{\bmP+\bmk} &&~~\otimes ~~~~~~\ket{\bmp} &&\otimes ~~~~\ket{\bmk}\,, 
    \label{three-body_state_free}
    \\
    \ket{\hat{\beta}\psi^{-}_{12k}}&\eqcl \sqrt{8E_1E_2 \omega}
    &&\underbrace{\ket{\bmP+\bmk}}_{\text{3-particle COM}} 
    &&~~\otimes 
    \underbrace{\ket{\Psi^{-}_{\bmp}}}_{\substack{\text{2-body} \\ \text{relative motion}}} 
    &&\otimes ~~
    \underbrace{\ket{\Psi^{-}_{\bmk}}}_{\text{graviton}}
    \,.
    \label{three-body_state}
\end{alignat}
\end{tcolorbox}
\noindent
There are a few comments. The second factorisation is expected to hold in the classical limit, where we have a large hierarchy between the massive and massless momenta, $p_a,p_a'\gg k$ (we use ``$\eqcl$'' for equalities under \eqref{three-body_state} to keep track of using this factorisation assumption). Nevertheless, the graviton momentum $\bmk$ should {\it not} be neglected in the COM state of the three bodies $\ket{\bmP+\bmk}$, as we will see shortly and in Sec.~\ref{sec:waveform}. Next, since the two-body potential of $\ket{\hat{\beta}\psi^{-}_{12k}}$ coincides with the conjugate of the two-body potential $\hat{V}_{\rm PM}$, the relative state of 1 and 2 is the same as the conjugate of the one in \eqref{two-body_state}. Finally, we retain the general scattering state $\ket{\Psi^{\pm}_{\bmk}}$ to see the structures of amplitudes and observables, although we have not derived the Schr\"{o}dinger equation for this. When necessary, we shall simply approximate it to be a free state $\ket{\Psi^-_{\bmk}}=\ket{\bmk}$.

All in all, our problem is to find the interacting state of the relative motion states $\ket{\Psi^{\pm}_{\bmp}}$. Using \eqref{two-body_state}, eq.~\eqref{PM_potential} is rewritten as
\begin{align}
    \braket{\bmP';\bmp'|\hat{V}_{\rm PM}|\bmP;\bmp}=\hdelta^{(3)}(\bmP'-\bmP)V_{\rm PM}(\bmp',\bmp;\bmP)
    \,,
\end{align}
and the Schr\"{o}dinger equation \eqref{Sch_alpha} yields
\begin{align}
    (E_{12}-E_{12}')\braket{\bmp'|\Psi^{+}_{\bm{p}}}-\int \hd^3 \bmr \, V_{\rm PM}(\bmp',\bmr;\bmP)\braket{\bmr|\Psi^{+}_{\bm{p}}}=0
    \,.
    \label{Sch_momentum}
\end{align}
Here, the potential can depend on the total momentum $\bmP$ because we have not assumed the COM frame. Note that the Schr\"{o}dinger equation for the out state reads
\begin{align}
(E_{12}-E_{12}')\braket{\bmp'|\Psi^{-}_{\bm{p}}}-\int \hd^3 \bmr\, V^*_{\rm PM}(\bmr,\bmp';\bmP)\braket{\bmr|\Psi^{-}_{\bm{p}}}=0
    \,,
\end{align}
with the different boundary condition, but the way to find the solution is formally the same as the in state (and only the in state is needed to compute the waveform in the end). Therefore, we only consider the in state \eqref{Sch_momentum} in the following.

Let us move to the position space. The position-space potential is defined by the Fourier transform~\cite{Neill:2013wsa}
\begin{align}
    V_{\rm PM}(\bar{\bm{p}},\bm{x};\bmP) &:= \int \hd^3 \bm{q}\, e^{-i\bm{q}\cdot \bm{x}}V_{\rm PM}(\bm{p}',\bm{p};\bmP)|_{\bmp=\bar{\bmp}+\frac{1}{2}\bmq,\bmp'=\bar{\bmp}-\frac{1}{2}\bmq}\,,
    \label{WFourier}
\end{align}
with $\bm{q}:=\bm{p}-\bm{p}'$ being the momentum transfer and $\bar{\bmp}:=\frac{1}{2}(\bmp+\bmp')$.\footnote{With this definition, the position-space potential becomes real when it is Hermitian $\hat{V}_{\rm PM}=\hat{V}_{\rm PM}^{\dagger}$. In Section~\ref{sec:waveform}, we will assume a Hermitian potential, but we do not make that assumption in this section.} The position-space Schr\"{o}dinger equation is obtained by performing the Fourier transform of \eqref{Sch_momentum} with respect to $\bm{p}'$. Notice that
\begin{align}
    &\int \hd^3 \bmp' \hd^3 \bmr\, V_{\rm PM}(\bmp',\bmr; \bmP) e^{i\bmp'\cdot \bmx} \braket{\bmr|\Psi^{+}_{\bm{p}}}
    \nn
    &=\int \hd^3 \bmp' \hd^3 \bmr \dd^3 \bmx'\, V_{\rm PM}(\bmp',\bmr; \bmP) e^{i\bmp'\cdot \bmx-i\bmr \cdot \bmx'} \Psi^{+}_{\bm{p}}(\bmx')
    \nn
    &=\int \hd^3 \bar{\bmp} \dd^3 \bmx' e^{-i\bar{\bmp}\cdot (\bmx'-\bmx)} \Psi^{+}_{\bm{p}}(\bmx')\int \hd ^3\bmq \,e^{-i \bmq \cdot \bmx'+\frac{i}{2}\bmq\cdot (\bmx'-\bmx)} V_{\rm PM}(\bar{\bmp},\bmq; \bmP) 
    \label{WFourier2}
\end{align}
where $\Psi^+_{\bm p}(\bm{x})=\braket{\bmx|\Psi^+_{\bmp}}$ is the wavefunction in the position space. To write this integral in terms of \eqref{WFourier}, we expand $e^{\frac{i}{2}\bmq \cdot (\bmx'-\bmx)}$ and rewrite the series by using the derivatives:
\begin{align}
    \eqref{WFourier2} &= \int \hd^3 \bar{\bmp} \dd^3 \bmx' e^{-i\bar{\bmp}\cdot (\bmx'-\bmx)} \Psi^{+}_{\bm{p}}(\bmx')\int \hd ^3\bmq e^{-i \bmq \cdot \bmx'}\left[1 +\frac{i}{2}\bmq\cdot (\bmx'-\bmx)+\cdots \right] V_{\rm PM}(\bar{\bmp},\bmq; \bmP) 
    \nn
    &=\int \hd^3 \bar{\bmp} \dd^3 \bmx' e^{-i\bar{\bmp}\cdot (\bmx'-\bmx)} \Psi^{+}_{\bm{p}}(\bmx') \left[1 - \frac{1}{2}(\bmx'-\bmx)\cdot \partial_{\bmx'} +\cdots \right] V_{\rm PM}(\bar{\bmp},\bmx';\bmP)
    \nn
    &=\int \hd^3 \bar{\bmp} \dd^3 \bmx' e^{-i\bar{\bmp}\cdot (\bmx'-\bmx)} \Psi^{+}_{\bm{p}}(\bmx') \left[1 + \frac{i}{2}\partial_{\bar{\bmp}}\cdot \partial_{\bmx'} +\cdots \right] V_{\rm PM}(\bar{\bmp},\bmx';\bmP)
    \,.
\end{align}
We have replaced $ e^{-i\bar{\bmp}\cdot (\bmx'-\bmx)}(\bmx'-\bmx)$ with $i\partial_{\bar{\bmp}} e^{-i\bar{\bmp}\cdot (\bmx'-\bmx)}$ and then performed the integration by parts to obtain the last equation. Therefore, the Schr\"{o}dinger equation in the position space is given by, after relabelling the integration variable $\bar{\bmp}\to \bmp'$,
\begin{align}
    E_{12}\Psi^{+}_{\bm{p}}(\bm{x})-\int \dd^3 \bm{x}' \hd^3\bmp'\,   e^{-i \bmp' \cdot (\bm{x}'-\bm{x})}\Psi^{+}_{\bm{p}}(\bm{x}')\left[1 + \frac{i}{2}\partial_{\bmp'}\cdot \partial_{\bmx'} +\cdots \right] H_{12}(\bmp',\bmx';\bmP)=0\,.
    \label{Sch_position}
\end{align}
with the Hamiltonian
\begin{align}
    H_{12}(\bmp,\bmx;\bmP)=\sqrt{m_1^2+\bmp_1^2}+\sqrt{m_2^2+\bmp_2^2}+V_{\rm PM}(\bmp,\bmx;\bmP)
    \,.
\end{align}
From \eqref{Sch_position}, one may obtain the familiar differential form of the Schr\"{o}dinger equation by replacing the $\bm{p}'$-dependence of the Hamiltonian according to $\bm{p}' \to \hat{\bm{p}}'=-i \partial_{\bmx'}$ and performing the $\bmx'$ and $\bm{p}'$ integrals (see the end of Appendix~\ref{sec:saddle}). We, however, work in the integral form \eqref{Sch_position} because the Hamiltonian $H_{12}$ is a non-linear function of $\bmp$ and thus the differential form is not practically useful. 

We solve the Schr\"{o}dinger equation \eqref{Sch_position} under the WKB approximation. Reintroducing $\hbar$, eq.~\eqref{Sch_position} becomes
\begin{align}
     E_{12}\Psi^{+}_{\bm{p}}(\bm{x})-\int \frac{\dd^3 \bm{x}' \dd^3\bmp'}{(2\pi \hbar)^3}\,   e^{-i \bmp' \cdot (\bm{x}'-\bm{x})/\hbar}\Psi^{+}_{\bm{p}}(\bm{x}')\left[1 + \frac{i\hbar}{2}\partial_{\bmp'}\cdot \partial_{\bmx'} +\cdots \right] H_{12}(\bmp',\bmx';\bmP)=0\,.
    \label{Sch_hbar}
\end{align}
Let us consider the WKB ansatz
\begin{align}
    \Psi^+_{\bmp}(\bm{x})=A_{\bmp}(\bm{x})e^{\frac{i}{\hbar}\sigma_{\bmp}(\bm{x})}
    \,, 
\end{align}
and evaluate the integral in \eqref{Sch_hbar} around the saddle
\begin{align}
    \bm{x}=\bm{x}'\,, \qquad \bm{p}'=\frac{\partial \sigma_{\bmp}(\bmx')}{\partial \bm{x}'}  =: \bmpb(\bmx')
    \,.
    \label{saddle}
\end{align}
The saddle-point approximation up to the sub-leading order yields
\begin{align}
    E_{12}(\bmp;\bmP)-H_{12}(\bmpb(\bmx),\bm{x};\bmP)&=0
    \,, \label{eq_bmp} \\
    \partial_{\bmx}\cdot\left[ \bm{v}_{\rm bulk}(\bmx) A_{\bmp}^2(\bmx) \right]&=0
    \,, \label{eq_continuity}
\end{align}
with $\bm{v}_{\rm bulk}:=\left. \frac{\partial H_{12}}{\partial \bmp}\right|_{\bmp=\bmp_{\rm bulk}(\bmx)}$. Solving the second equation under the boundary condition that $A_{\bmp}\to 1$ as the free wave limit $\sigma_{\bmp}\to\bmp \cdot \bmx$, we find\footnote{\label{footnote}We assume that $\sigma_{\bmp}$ satisfies the boundary condition for scattering in states. If one is interested in the emission from a bound state, one should instead use the bound state wavefunction by changing the boundary condition. }
\tbox{WKB wavefunction}{
    \Psi^+_{\bmp}(\bmx)=\sqrt{\det \partial_{\bmp} \partial_{\bmx} \sigma_{\bmp}(\bmx)}\, e^{\frac{i}{\hbar}\sigma_{\bmp}(\bmx)}\,.
    \label{WKB_wf2}
}
The detailed derivation of \eqref{eq_bmp}-\eqref{WKB_wf2} is described in Appendix~\ref{sec:saddle}. Since \eqref{eq_bmp} is the classical Hamilton-Jacobi equation in the {\it bulk}, the quantity $\bmpb(\bmx)=\partial_{\bmx}\sigma_{\bmp}(\bmx)$ in \eqref{eq_bmp} can interpreted as the {\it bulk} momentum in the classical limit (and $\sigma_{\bmp}$ is Hamilton's characteristic function).\footnote{However, this interpretation requires care because our potential is not necessarily real. An imaginary part of the two-body potential $V_{\rm PM}$, even if it is $\hbar$ suppressed, will lead to an imaginary part of $\sigma_{\bmp}$, which changes the overall size of the wavefunction. }

Note that \eqref{eq_bmp}-\eqref{WKB_wf2} arise from a saddle-point approximation applied to a Hamiltonian system with a fully relativistic kinetic term and potential. It has been shown that such a system can be recast into a non-relativistic–looking form by quantising, rather than the Hamiltonian itself, the solution in phase space. This observation was first made at leading order in \cite{Damour:2017zjx}, later conjectured to hold to all orders in \cite{Bern:2019crd}, and subsequently proven in \cite{Kalin:2019rwq,Bjerrum-Bohr:2019kec}. One advantage of this approach is that $\bmpb^2(\bmx)$ in isotropic coordinates is proportional to the finite part of a 2-to-2 amplitude to all orders.\footnote{This relation holds only in $D=4$, as it already fails at one loop for $D>4$. For a proof, see Section 4 of \cite{Cristofoli:2020uzmZ}.} Here, instead, we returned to the relativistic form to determine the wavefunction in generic coordinates, including the normalisation $A_{\bmp}(\bmx)$, which is crucial during wave generation.

The analysis so far has not assumed any specific coordinates. We wish to assume the COM frame to simplify \eqref{eq_bmp}. However, even if we take the COM frame initially $\bmP=0$, the state evolves into the three-particle state with the momentum conservation $\bmP'+\bmk=\bmP$, meaning that the two bodies cannot remain in the COM frame. As for massless particles, what we keep finite in the classical limit is the wavenumber $\bar{\bmk}=\bmk/\hbar$, so the deviation from the COM frame is $\hbar$-suppressed. Let us thus solve \eqref{eq_bmp} with $\bmP=\mathcal{O}(\hbar)$ (or the corresponding equation for the out state). The solutions should be given by
\begin{align}
    \sigma_{\bmp}=\sigma^{\rm COM}_{\bmp} + \hbar\, \delta \sigma_{\bmp}
\end{align}
where $\sigma_{\bmp}^{\rm COM}$ is the solutions in the COM frame $\bmP=0$. Looking at the WKB wavefunction \eqref{WKB_wf2}, one sees that the correction to the phase $\delta \sigma_{\bmp}$ will give a {\it finite} contribution even in the classical limit $\hbar\to 0$. This is precisely the reason why we must be cautious of deviations from the COM frame. In Sec.~\ref{sec:waveform}, we will see that $\delta \sigma_{\bmp}(\bmx)$ is indeed necessary to reproduce the classical wave emission from worldlines.

We now fix the definition of the COM. A convenient choice is
\begin{align}
    \eta_1=\frac{E_1}{E_{12}}\,, \qquad \eta_2=\frac{E_2}{E_{12}}
    \label{eta12}
\end{align}
where the COM position $\bmX$ conjugate to the total momentum remains constant in the COM frame when the interaction is turned off, i.e., $E_1(\bmp_1)+E_2(\bmp_2)=E_1(\bmp)+E_2(-\bmp)+\mathcal{O}(\bmP^2)$. The equation for $\delta \sigma_{\bmp}$ is then found to be
\begin{align}
     \frac{\partial\, \delta \sigma_{\bmp}}{\partial \bmx} \cdot \left.\frac{\partial H_{12}}{\partial \bmpb}\right|_{\bmP=0} + \bmVb \cdot \bm{P} =0\,,
     \label{eq_delta_sigma}
\end{align}
where
\begin{align}
    \bmVb(\bmx):=\left. \frac{\partial H_{12}}{\partial \bmP} \right|_{\bmP=0} = \left. \frac{\partial V_{\rm PM}}{\partial \bmP}\right|_{\bmP=0}
\end{align}
is understood as the bulk velocity of the COM position $\bmX$ in the classical limit.\footnote{The COM frame $\bmP=0$ does not necessarily mean $\bmVb=0$ in relativistic problems.} We leave a further discussion of \eqref{eq_delta_sigma} until Sec.~\ref{sec:waveform}.

\subsection{Wavefunctions to resummed amplitudes}
\label{sec:wf_to_amp}
We finally compute the $T$-matrix by using the radiation potential \eqref{R_potential} and non-perturbative wavefunctions \eqref{WKB_wf2}. We come back to the natural units $\hbar=1$ to simplify the notation. We use the factorisations \eqref{two-body_state} and \eqref{three-body_state} and work in the COM frame of the initial state $\bmP=0$ for simplicity. The $T$-matrix of the one-graviton emission process is then written by
\begin{align}
T_{\beta,\alpha}&\eqcl \sqrt{32E_1 E_2 E'_1 E'_2 \omega} \braket{\Psi^-_{\bmp'};\Psi^-_{\bmk} |\hat{R}_{\rm PM}|\Psi^+_{\bmp}}\hdelta^{(3)}(\bmP'+ \bmk)
\,.
\label{5pt_T-mat}
\end{align}
The kernel of the $T$-matrix is thus the following integrals
\begin{align}
    \braket{\Psi^-_{\bmp'};\Psi^-_{\bmk}|\hat{R}_{\rm PM}|\Psi^+_{\bmp}}
    &=\int \hd^3 \bml \braket{\Psi^-_{\bmk}|\bml} \, \mathcal{R}
    \label{k_int}
\end{align}
and
\begin{align}
    \mathcal{R}&:= \int \hd^3 \bmr \, \hd^3 \bmr' \dd^3 \bmx \dd^3 \bmx' 
    [\Psi^-_{\bmp'}(\bmx')]^* e^{i\bmr' \cdot \bmx'}R_{\rm PM}(\bmr',\bml,\bmr )  e^{-i\bmr \cdot \bmx} \Psi^+_{\bmp}(\bmx)
    \,.
    \label{R_int}
\end{align}
Here, $R_{\rm PM}$ is the momentum-space radiation potential originally introduced in \eqref{R_potential}. Using \eqref{two-body_state} and \eqref{three-body_state_free} with the change of notation $\bmp \to\bmr$ and $ \bmk\to\bml$, it is given by
\begin{align}
    \braket{\bmP'+\bml;\bmr';\bml|\hat{R}_{\rm PM}|\bmP=\bm{0};\bmr} = \hdelta(\bmP'+\bml)R_{\rm PM}(\bmr',\bml,\bmr)\,.
\end{align}

We define the position-space radiation potential by\footnote{We define it for the outgoing momentum $\bmr'$. Instead, one can define it for the average momentum $\frac{1}{2}(\bmr'+\bmr)$ to keep the symmetry between the in and out momenta, and follow similar steps to \eqref{WFourier2}.}
\begin{align}
    R_{\rm PM}(\bmr',\bml,\bmx):=\int \hd^3 \bmq \,e^{-i\bmq\cdot \bmx}R_{\rm PM}(\bmr',\bml,\bmr)|_{\bmr=\bmr'+\bmq}
    \label{def_R_position}
\end{align}
so that the integral \eqref{R_int} can be written as
\begin{align}
    \mathcal{R} &=\int \hd^3 \bmr' \dd^3 \bmx \dd^3 \bmx' [\Psi^-_{\bmp'}(\bmx')]^*e^{i\bmr'\cdot (\bmx'-\bmx)}R_{\rm PM}(\bmr',\bml,\bmx)  \Psi^+_{\bmp}(\bmx)
    \,.
    \label{R_int2}
\end{align}
We assume that the position-space radiation potential has no super-classical part, that is, it is regular in the classical limit $\hbar\to 0$.
The integrals of $\bmr'$ and $\bmx$ can then be performed by applying the saddle-point approximation. The rapidly oscillating parts of \eqref{R_int2} come from $e^{i\bmr'\cdot (\bmx'-\bmx)}$ and $\Psi^+_{\bmp}(\bmx) \propto e^{i\sigma^{\rm COM}_{\bmp}(\bmx)}$. Therefore, the saddle is
\begin{align}
    \bmx=\bmx'\,, \qquad \bmr'=\partial_{\bmx}\sigma^{\rm COM}_{\bmp}(\bmx)=\bmpb(\bmx)
    \,.
\end{align}
By relabelling $\bmx'\to \bmx$ to simplify the notation, we finally obtain
\tbox{5-point amplitude kernel}{
   \mathcal{R} &= \int\dd^3 \bmx [\Psi^-_{\bmp'}(\bmx)]^* R_{\rm PM}(\bmpb(\bmx),\bml,\bmx)  \Psi^+_{\bmp}(\bmx)
   \,.
   \label{R_int_f}
}
Eqs.~\eqref{5pt_T-mat}, \eqref{k_int}, and \eqref{R_int_f} are our general integral expressions to compute the non-perturbative 5-point amplitude under the WKB approximation. They are valid for the generic two-body potential and radiation potential to all orders in $G$ under the WKB approximation and the factorisation approximation of the three-particle state \eqref{three-body_state}.

We conclude this section by providing the explicit form of the position-space radiation potential. In general, classically relevant contributions of $R_{\rm PM}(\bmp',\bmk,\bmx)$ are expected to arise from singularities of the momentum-space potential in the momentum transfer $\bmq$. At the leading order \eqref{Rmomentum_LO}, the singularities are the delta functions. In the classical limit, the outgoing momenta are approximated by
\begin{align}
    \bmp_1'=\bmp'+\eta_1(\bmp_1',\bmp_2') \bmP' \simeq \bmp'-\eta_1(\bmp'{}^2) \bmk\,, ~~
    \bmp_2'=-\bmp'+\eta_2(\bmp_1',\bmp_2') \bmP' \simeq -\bmp'-\eta_2(\bmp'{}^2)\bmk\,,
\end{align}
where we recall that the two-body out state is no longer in the COM frame due to the momentum conservation $\bmP'+\bmk=\bmP=0$. Hence, the momentum transfer of each particle is
\begin{align}
    \bmq_1:=\bmp_1-\bmp_1'\simeq \bmq+\eta_1(\bmp'{}^2)\bmk\,, \qquad \bmq_2:=\bmp_2-\bmp_2'\simeq -\bmq+\eta_2(\bmp'{}^2)\bmk
    \,.
\end{align}
Then, the $\bmq$ integral in \eqref{def_R_position} is trivial thanks to the delta functions in \eqref{Rmomentum_LO}, yielding
\tbox{Leading-order radiation potential in position space}{
    R_{\rm PM}(\bmp',\bmk,\bmx)= \frac{\kappa}{\sqrt{2\omega}}\left[e^{-i\bmk \cdot \bar{\bmx}_1} \frac{(\varepsilon \cdot p'_1)^2}{2E_1(\bmp'{}^2)}+e^{-i\bmk\cdot \bar{\bmx}_2} \frac{ (\varepsilon \cdot p'_2)^2}{2E_2(\bmp'{}^2)}\right] + \mathcal{O}(\kappa^3)
    \label{RLO_position}
}
in the classical limit. The denominator of each term has been simplified by $E_a \simeq E_a'$, and the barred coordinates
\begin{align}
    \bar{\bmx}_1=\eta_2(\bmp'{}^2) \bmx\,, \qquad \bar{\bmx}_2=-\eta_1(\bmp'{}^2) \bmx
\end{align}
are the relative coordinates of particle $a$ from the COM position $\bmX$.


\section{Waveform from generic conservative dynamics}
\label{sec:waveform}

In Sec.~\ref{sec:resum}, we have developed a resummation of the 5-point amplitude by applying the twice-projection of Feshbach. The framework is general and does not require specific assumptions about the potentials as long as the WKB approximation and the factorisation approximation are valid. From now on, we ignore the imaginary parts in the potentials, especially the one in the two-body potential $V_{\rm PM}$, and focus on GW emissions from the conservative motion of two bodies. More precisely, we approximate that the two-body potential is Hermitian $\hat{V}_{\rm PM}=\hat{V}_{\rm PM}^{\dagger}$, which reads 
\begin{align}
    V_{\rm PM}(\bmp',\bmp;\bmP)=V^*_{\rm PM}(\bmp,\bmp';\bmP)
\end{align}
in the momentum space, and
\begin{align}
    V_{\rm PM}(\bar{\bmp},\bmx;\bmP)=V^*_{\rm PM}(\bar{\bmp},\bmx;\bmP)
\end{align}
in the position space. Nonetheless, except for this additional approximation, the following formulae remain all orders in $G$ and are applicable to generic conservative dynamics.

\subsection{Waveform as in-in expectation value}
\label{sec:KMOCwaveform}
\begin{figure}[t]
\centering
\includegraphics[width=0.7\linewidth]{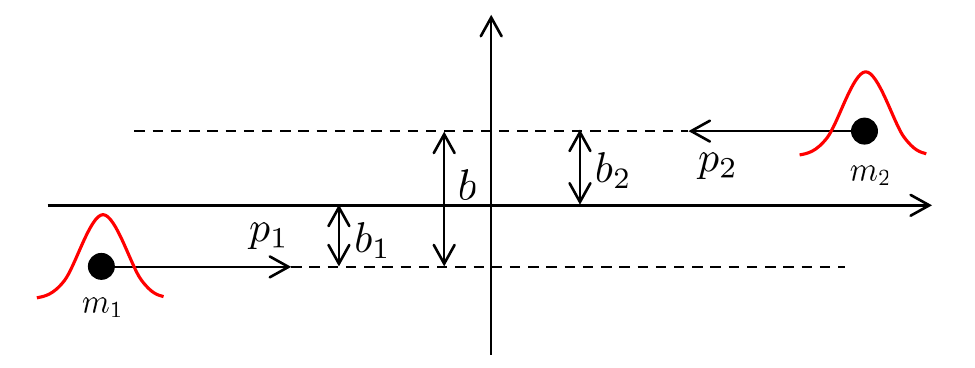}
\caption{The initial kinematics of the KMOC formalism.}
\label{fig:KMOC}
\end{figure}

We calculate gravitational waveforms by using the KMOC formalism~\cite{Kosower:2018adc, Cristofoli:2021vyo}. To deal with the classical two-body problem, the initial state is chosen to be a wavepacket state so that both position and momentum are well-defined in the classical limit; in the interaction picture, our initial state is
\begin{align}
    \ket{\Psi}:=\int d\Phi(p_1,p_2)\phi_1(p_1)\phi_2(p_2)e^{i(b_1\cdot p_1+b_2 \cdot p_2)}\ket{p_1; p_2}
    \,.
\end{align}
The wavefunctions $\phi_a$ are normalized to unity
\begin{align}
    \int \dd \Phi(p_a)|\phi_a(p_a)|^2 = 1
    \,,
\end{align}
and assumed to be sharply peaked at a classical value of the momentum of the body. They are shifted by the translation operators $e^{i(b_1\cdot p_1+b_2 \cdot p_2)}$ with spatial vectors $b_a$, and
\begin{align}
    b:=b_2-b_1
\end{align}
is the impact parameter. See Fig.~\ref{fig:KMOC} for illustration.

Classical observables at future infinity are expectation values of operators after time evolution governed by the $S$-matrix operator $\hat{S}$. The waveform is the expectation value of the graviton operator at future null infinity, spanned by the retarded time $u$ and the unit vector $\bm{n}$ directing to a point on the sphere, so given by~\cite{Cristofoli:2021vyo}
\begin{equation}
    h_{\mu\nu}(u,\bm{n})=\frac{\kappa}{4\pi r} \sum_{\sigma} \int \hd \omega\, \varepsilon^{-\sigma}_{\mu\nu} e^{-i\omega u} W^{\sigma}(k)|_{k=(\omega,\,\omega \bm{n})}
    +{\rm c.c.},
\end{equation}
with
\begin{align}
iW^{\sigma}(k)&:=\braket{\Psi|\hat{S}^{\dagger}\hat{a}_k^{\sigma}\hat{S}|\Psi} \,,
\end{align}
being the spectral waveform. Here, $a^{\sigma}_k$ is the annihilation operator of a graviton with the polarisation $\sigma=\pm$ and $\varepsilon_{\mu\nu}^{\sigma}(k)$ is the polarisation tensor. In what follows, the polarisation $\sigma$ is implicit as it is insignificant. Under the appropriate choice of the wavefunction satisfying the so-called Goldilocks condition, the spectral waveform in the classical limit is
\begin{align}
iW(k)&=
\int\left[ \prod_{a=1,2} \hd^4 q_{a} \hdelta(2p_{a} \cdot q_{a})e^{ib_a \cdot q_{a}} \right] \braket{p'_{1};p'_{2}|\hat{S}^{\dagger}\hat{a}_{k}\hat{S}|p_{1};p_{2}}
\,,
\end{align}
where $q_{a}$ is the momentum mismatch of the plane waves composed of the initial wavepacket state $ q_{a}:=p_a-p'_{a}$ and $p_a$ are identified with the classical on-shell momenta of the initial two bodies. The $q_a$ integrals arise from the initial state being the wavepacket, whose domain is restricted to be the on-shell momenta of $p_a'=p_a-q_a$ as guaranteed by the delta functions, and to be $q_a=\mathcal{O}(\hbar)$ because the wavepacket must be supported only around the classical momentum $p_a$.

Note that, with $p_a^2=m_a^2$, the on-shell condition of $p_a'$ is written as $2p_a\cdot q_a-q_a^2=0$. We have ignored $q_a^2$ by the scaling $q_a=\mathcal{O}(\hbar)$. Alternatively, one can relate the barred variable $\bar{p}_a=(p_a+p_a')/2$ to the classical four-velocity $u_a$ via $\bar{p}_a=\bar{m}_a u_a$ with $\bar{m}_a^2=m_a^2-q_a^2/4$ to keep the exact form of the on-shell condition $2\bar{p}_a\cdot q_a=0$ as in \cite{Brandhuber:2023hhy,Herderschee:2023fxh}. This subtlety does not enter our computation because the super-classical parts are manifestly cancelled in the waveform computations, thanks to the resummed form of the wavefunction. We also compute the impulse in Appendix~\ref{sec:impulse}, and show that a non-perturbative impulse can be obtained even if $q_a^2$ is ignored.

Therefore, in the KMOC formalism, the kernel of the waveform is the inclusive amplitude
\begin{align}
\braket{p'_{1};p'_{2}|\hat{S}^{\dagger}\hat{a}_{k}\hat{S}|p_{1};p_{2}}
    \,.
    \label{inc_ak}
\end{align}
The inclusive amplitude \eqref{inc_ak} can be rewritten in terms of in-out amplitudes by inserting the completeness relation. In our context, the conservative dynamics refers to discarding the imaginary part of the potentials. The imaginary part of the two-body potential should arise from cutting the amplitudes in terms of on-shell inelastic states. Hence, at the level of conservative dynamics, the relevant on-shell contributions would be the elastic states only, which motivate us to truncate the completeness relation up to the elastic channel:\footnote{We also ignore bound states in the completeness relation because in the classical limit the scattering state does not evolve into a bound state at the conservative level.}
\begin{align}
    \hat{1}\eqcons \int \dd \Phi(p_{1},p_{2})  \ket{p_{1};p_{2}}\bra{p_{1};p_{2}} 
    \,,
    \label{completness_cons}
\end{align}
where $\eqcons$ denotes equalities under our assumption on the conservative dynamics.
Then, the inclusive amplitude is
\begin{align}
\braket{p'_{1};p'_{2}|\hat{S}^{\dagger}\hat{a}_{k}\hat{S}|p_{1};p_{2}}
&\eqcons \int \dd \Phi(p''_{1},p''_{2}) \braket{p'_{1};p'_{2}|\hat{S}^{\dagger}|p''_{1};p''_{2}}\braket{{p''_{1};p''_{2}|\hat{a}_{k}\hat{S}|p_{1};p_{2}}}
\nn
&\,=\int \dd \Phi(p''_{1},p''_{2}) \braket{p'_{1};p'_{2}|\hat{S}^{\dagger}|p''_{1};p''_{2}}
\braket{{p''_{1};p''_{2};k|\hat{S}|p_{1};p_{2}}}
\,.
    \label{inc_ak_con}
\end{align}

Let us use the results from Sec.~\ref{sec:resum}. Using \eqref{two-body_state} and \eqref{three-body_state}, the 2-to-2 and 2-to-3 $S$-matrix elements are written as
\begin{align}
    \braket{p'_{1};p'_{2}|\hat{S}^{\dagger}|p_{1};p_{2}}
    &= \sqrt{16E'_{1} E'_{2} E_{1} E_{2}}\,\hdelta^{(3)}(\bmP'-\bmP)\braket{\Psi^+_{\bmp'}|\Psi^-_{\bmp}}
    \,, 
    \label{S22_relative} \\
    \braket{{p'_{1};p'_{2};k|\hat{S}|p_{1};p_{2}}}
    &\eqcl -i\sqrt{32E_{1} E_{2} E'_{1} E'_{2} \omega}\, \hdelta^{(4)}(P'+k-P) \braket{\Psi^-_{\bmp'};\Psi^-_{\bmk}|\hat{R}_{\rm PM}|\Psi^+_{\bmp}}
    \,,
    \label{S23_relative}
\end{align}
with $P^{\mu}=(E_{12},\bmP)$ and $k^{\mu}=(\omega,\bmk)$. Here, we write the 2-to-2 $S$-matrix in the form of the inner product of in and out states. In \eqref{S23_relative}, we combined the energy delta function in \eqref{S-matrix} with the momentum delta function in \eqref{5pt_T-mat}. We then use the fact that the interacting states also satisfy the completeness relation for the conservative dynamics
\begin{align}
   \hat{1}\eqcons \int \hd^3 \bmp \ket{\Psi^-_{\bmp}}\bra{\Psi^-_{\bmp}} 
   \,.
   \label{completness_sc}
\end{align}
As a result, after performing the total momentum integral, the inclusive amplitude under both the classical limit and the conservative dynamics is given by
\tbox{In-in formula of inclusive amplitude}{
\braket{p'_{1};p'_{2}|\hat{S}^{\dagger}\hat{a}_{k}\hat{S}|p_{1};p_{2}}
    &\eqcc -i\sqrt{32E_{1} E_{2} E'_{1} E'_{2} \omega}\, \hdelta^{(4)}(P'+k-P) \braket{\Psi^+_{\bmp'};\Psi^-_{\bmk}|\hat{R}_{\rm PM}|\Psi^+_{\bmp}}
    \,.
    \label{incl_in-in}
}
This expression is quite close to that of the 5-point $S$-matrix \eqref{S23_relative}, but the out state of the two bodies is replaced with the in state. In fact, Refs.~\cite{Caron-Huot:2023vxl, Caron-Huot:2023ikn} pointed out that the inclusive amplitude is related to the 5-point amplitude via analytic continuation, which changes the sign of $i\varepsilon$ of the out two bodies to opposite. Our formulae \eqref{S23_relative} and \eqref{incl_in-in} indeed suggest such a structure as they are related by $\Psi^-_{\bmp'} \leftrightarrow \Psi^+_{\bmp'}$. Note also that the expression analogous to \eqref{incl_in-in} was obtained in the external mass-ratio limit~\cite{Aoki:2024boe}~(see also~\cite{Adamo:2023cfp} for the earlier development). In this case, $\Psi^{\pm}_{\bmp}$ and $\Psi^-_{\bmk}$ are given by the solutions to the Klein-Gordon equations in the Schwarzschild background under the in/out boundary conditions. However, we stress that the formula~\eqref{incl_in-in} is derived under generic assumptions of the classical limit and the conservative dynamics to all orders in $G$, without any assumption of the mass ratio.

It would be worth mentioning at this stage how the analysis above should be corrected in non-conservative dynamics. First of all, we should include additional contributions in the completeness relation \eqref{completness_cons}. In fact, it is well-known that the computation of the impulse with the radiation reaction requires additional cut contributions in the KMOC formalism~\cite{Kosower:2018adc}. Second, a non-Hermitian system is biorthogonal, requiring dual states to form a complete set~\cite{Muga:2004zz}. Therefore, \eqref{completness_sc} does not hold, invalidating to rewrite of the inclusive amplitude into the simple in-in form \eqref{incl_in-in}. Having said that, the results in Sec.~\ref{sec:resum} hold even in non-conservative systems with non-Hermitian potentials. It would be interesting to get a more systematic understanding of the non-conservative case, which we leave for future study.

\subsection{Resummed waveform}

Collecting all our developments, let's compute the KMOC waveform:
\begin{align}
    W\eqcc -\int &\hd^4 q_{1} \hd^4 q_{2} \hdelta(2p_1 \cdot q_{1}) \hdelta(2p_{2} \cdot q_{2}) \hdelta^{(4)}(q_{1}+q_{2}-k) e^{i(b_1 \cdot q_{1} + b_2 \cdot q_{2})}
\nn
&\times 4E_{1}E_{2}   \sqrt{2\omega} \braket{\Psi^+_{\bmp-\bmq};\Psi^-_{\bmk}|\hat{R}_{\rm PM}|\Psi^+_{\bmp}}
\,,
\end{align}
where we have simplified the overall factor of \eqref{incl_in-in} under the classical limit $E_{a}\simeq E'_{a}$ and rewritten the delta functions for the four-momentum conservation in terms of the momentum mismatches $q_{a}=p_a-p'_{a}=\mathcal{O}(\hbar)$.
We introduce the following variables
\begin{align}
    p_{1}&=\eta_{1} P+p\,, \qquad p_{2}=\eta_{2} P-p
    \,, \label{PQ_var1} \\
    q_{1}&=\eta_{1} Q+q\,, \qquad q_{2}=\eta_{2} Q-q
    \,, \label{PQ_var2}
\end{align}
and change the integration variables from $q_a$ to $q$ and $Q$. We work in the COM frame with the choice~\eqref{eta12}, where the total and relative four-momenta are 
\begin{align}
    P^{\mu}=(E_{12}(\bmp^2),\bm{0})\,, \qquad p^{\mu}=(0,\bmp)
    \,.
\end{align}
The vectors $b_a$ are purely spatial $b_a=(0,\bm{b}_a)$. We perform the $Q$ integral and the $q^0$ integral by using the delta functions of the conservation and one of the on-shell conditions, yielding
\begin{align}
    W \eqcc -\sqrt{2\omega} \int \hd^3 \bmq \,\hdelta \left(\omega-\partial_{\bmp}E_{12}(\bmp^2) \cdot \bm{q}\right) e^{-i\bm{b}\cdot \bm{q} - i\bmX_0 \cdot \bmk}  \braket{\Psi^+_{\bmp-\bmq};\Psi^-_{\bmk}|\hat{R}_{\rm PM}|\Psi^+_{\bmp}}
    \label{W_delta}
\end{align}
where 
\begin{align}
    \bm{b}:=\bm{b}_1-\bm{b}_2\,, \qquad \bmX_0=\eta_{1} \bm{b}_1+\eta_{2} \bm{b}_2
\end{align}
are the impact parameter and the initial position of the COM, respectively.

The inclusive amplitude \eqref{incl_in-in} is computed in the same manner as Sec.~\ref{sec:wf_to_amp}. The spectral waveform is then
\begin{align}
    W&\eqcc \int \hd^3 \bml \braket{\Psi^-_{\bmk}|\bml} \mathcal{T}(\bml)\,,
    \label{waveform_T}
    \\
    \mathcal{T}(\bml)&:=
    -\sqrt{2\omega }\int\hd^3\bm{q} \dd^3 \bmx \, \hdelta\left(\omega-\partial_{\bmp}E_{12}(\bmp^2) \cdot \bm{q}\right) e^{-i\bm{b}\cdot \bm{q} - i\bmX_0 \cdot \bmk}
    \nn
    &\qquad \quad\times [\Psi^+_{\bmp-\bmq}(\bmx)]^* R_{\rm PM}(\bmpb(\bmx),\bml,\bmx)  \Psi^+_{\bmp}(\bmx)
    \,.
\end{align}
Alternatively, by using the wavefunction of the graviton $\Psi^-_{\bmk}(\bm{y})=\braket{\bm{y}|\Psi^-_{\bmk}}$, we obtain
\tbox{Resummed waveform}{
    W \eqcc \int \dd^3 \bm{y} [\Psi^-_{\bmk}(\bm{y})]^* \mathcal{T}(\bm{y})\,, \quad
    \mathcal{T}(\bm{y})=\int \hd^3 \bml e^{-i\bml \cdot \bm{y}}\,\mathcal{T}(\bml)
    \,.
    \label{waveform_position}}
By comparing the classical computation of the waveform, the $\bm{y}$ integral in \eqref{waveform_position} can be understood as the integral of the product of the position-space Green's function and the source term $\mathcal{T}(\bm{y})$. See Sec.~4 of~\cite{Aoki:2024boe} for a comparison between classical and amplitude computations in the context of black hole perturbation. Therefore, the wavefunction $\Psi^-_{\bmk}(\bm{y})$ is expected to be responsible for the propagation effect, and the generation of GWs should be described by $\mathcal{T}$.

Let us proceed with calculating the source term $\mathcal{T}(\bml)$. Our framework is based on time-independent scattering theory, where there has been no time explicitly. The time naturally emerges from the remaining on-shell delta function:
\begin{align}
    \hdelta\left(\omega-\bm{q}\cdot \partial_{\bmp}E_{12}(\bmp^2)\right) = \int \dd t \, e^{i\omega t -i \partial_{\bmp}E_{12}(\bmp^2)\cdot \bm{q}\, t}
    \,.
\end{align}
Using the WKB wavefunction \eqref{WKB_wf2} with care of the deviation from the COM frame, we obtain
\begin{align}
    \mathcal{T}(\bml)&= -\sqrt{2\omega }\int\hd^3\bm{q} \dd^3 \bmx \dd t \, e^{i\omega t}  
    e^{-i\bmX_0 \cdot \bmk-i\delta \sigma_{\bmp}}
    e^{-i  (\bm{b}+\partial_{\bmp}E_{12}(\bmp^2)t- \partial_{\bmp}\sigma^{\rm COM}_{\bmp})\cdot \bm{q} }
    \nn
    &\qquad \qquad\times |\det \partial_{\bmp} \partial_{\bmx} \sigma^{\rm COM}_{\bmp}(\bmx)|\, R_{\rm PM}(\bmpb(\bmx),\bml,\bmx)
    \nn
    &=-\sqrt{2\omega }\int \dd^3 \bmx \dd t \, e^{i\omega t} 
     e^{-i\bmX_0 \cdot \bmk-i\delta \sigma_{\bmp}}
     \delta^{3}(\bm{b}+\partial_{\bmp}E_{12}(\bmp^2)t- \partial_{\bmp}\sigma^{\rm COM}_{\bmp}(\bmx)) 
    \nn
    &\qquad \qquad\times |\det \partial_{\bmp} \partial_{\bmx} \sigma^{\rm COM}_{\bmp}(\bmx)|\,R_{\rm PM}(\bmpb(\bmx),\bml,\bmx)
\end{align}
where we used $\bmq=\mathcal{O}(\hbar)$ and ignored terms of $\mathcal{O}(\bmq^2)$ in the exponent. The super-classical piece of the WKB wavefunction is cancelled thanks to the in-in form of the inclusive amplitude. We notice that the argument of the delta function,
\begin{align}
    \bm{b}+\partial_{\bmp}E_{12}(\bmp^2)t- \partial_{\bmp}\sigma^{\rm COM}_{\bmp}(\bmx)=0
    \,,
\end{align}
is precisely the equation for the classical trajectory in the Hamilton-Jacobi formalism. Hence, denoting its solution by $\bmx=\bm{x}_{\rm cl}(t)$, we have
\begin{align}
    \delta^{(3)}(\bm{b}+\partial_{\bmp}E_{12}(\bmp^2)t- \partial_{\bmp}\sigma^{\rm COM}_{\bmp}(\bmx)) = \frac{\delta^{(3)}(\bmx-\bmx_{\rm cl}(t))}{|\det \partial_{\bmp} \partial_{\bmx} \sigma^{\rm COM}_{\bmp}(\bmx)|}
    \,,
\end{align}
and then
\begin{align}
    \mathcal{T}(\bml)&=-\sqrt{2\omega }\int \dd^3 \bmx \dd t \, e^{i\omega t} 
     e^{-i\bmX_0 \cdot \bmk-i\delta \sigma_{\bmp}}
     \delta^{(3)}(\bmx-\bmx_{\rm cl}(t))
     R_{\rm PM}(\bmpb(\bmx),\bml,\bmx)
    \,.
\end{align}
We finally use the support of the classical trajectory $\bmx=\bmx_{\rm cl}(t)$ to rewrite $\delta \sigma_{\bmp}$. We add the subscript ``cl'' when quantities are evaluated on the trajectory,
\begin{align}
    \bmp_{\rm cl}(t):=\bmpb(\bmx_{\rm cl}(t))\,,\qquad \bm{V}_{\rm cl}(t):=\bmVb(\bmx_{\rm cl}(t))
    \,,
\end{align}
to which we can apply Hamilton's equations. Eq.~\eqref{eq_delta_sigma} with the replacement $\bmP\to -\bmk$ can then be written by
\begin{align}
    \frac{\partial\, \delta \sigma_{\bmp}(\bmx_{\rm cl}(t))}{\partial \bmx_{\rm cl}} \cdot \frac{\dd \bmx_{\rm cl}}{\dd t} - \bm{V}_{\rm cl} \cdot \bm{k}=
    \frac{\dd\, \delta \sigma_{\bmp}(\bmx_{\rm cl}(t))}{\dd t} - \bm{V}_{\rm cl} \cdot \bm{k}=0
    \,.
\end{align}
We find
\begin{align}
    \bmX_0 \cdot \bmk+\delta \sigma_{\bmp}(\bmx_{\rm cl}(t)) = \left(\bmX_0+\int^t_{-\infty} \dd t' \bm{V}_{\rm cl}(t') \right) \cdot \bmk=\bmX_{\rm cl}(t)\cdot \bmk
\end{align}
with $\bmX_{\rm cl}(t)$ being the position of the COM at time $t$. All in all, our final formula is
\tbox{Source term of gravitational waves}{
    \mathcal{T}(\bml)=-\sqrt{2\omega}\int \dd t \  e^{i\omega t} 
     e^{-i\bmX_{\rm cl}(t)\cdot \bmk}
     R_{\rm PM}(\bmp_{\rm cl}(t),\bml,\bmx_{\rm cl}(t))
    \,.
    \label{T_formula}
}
This is our main result.

Let us use the leading order radiation potential \eqref{RLO_position} to see the waveform more concretely. Assuming the free propagation $\ket{\Psi^-_{\bmk}}=\ket{\bmk}$, we immediately find
\begin{align}
    W(k)&=\mathcal{T}(\bmk)
    \nn
    &=-\frac{\kappa}{2}\sum_{a=1,2}\int \dd t e^{i\omega t} e^{-i(\bar{\bmx}_{a,{\rm cl}}(t)+\bmX_{\rm cl}(t))\cdot \bm{k}} \frac{(\varepsilon \cdot p_{a,{\rm cl}}(t))^2}{E_{a,{\rm cl}}(t)}  +\mathcal{O}(\kappa^3)
    \nn
    &=-\frac{\kappa}{2}\sum_{a=1,2}\int \dd t e^{i\omega t} e^{-i\bmx_{a,{\rm cl}}(t)\cdot \bm{k}} \frac{(\varepsilon \cdot p_{a,{\rm cl}}(t))^2}{E_{a,{\rm cl}}(t)}  +\mathcal{O}(\kappa^3)
\end{align}
which is nothing but the Fourier transform of the energy-momentum tensors of worldlines
\begin{align}
    \int \dd t e^{i\omega t} e^{-i\bmx_{a,{\rm cl}}(t)\cdot \bm{k}} \frac{(\varepsilon \cdot p_{a,{\rm cl}}(t))^2}{E_{a,{\rm cl}}(t)} &= \varepsilon_{\mu\nu}(k) \int \dd^4 z e^{ik\cdot z}  \sum_{a=1,2}T_a^{\mu\nu}(z)
    \,, \\
   T^{\mu\nu}_a(z)&=\delta^{(3)}(\bm{z}-\bmx_{a,{\rm cl}}(t)) \frac{p^{\mu}_{a,{\rm cl}}p^{\nu}_{a,{\rm cl}}}{E_{a,{\rm cl}}}
   \,,
\end{align}
where $z^{\mu}=(t,\bm{z})$ and $\varepsilon^{\mu\nu}=\varepsilon^{\mu}\varepsilon^{\nu}$. One can see that the positions of the worldlines are correctly pulled back to the coordinate positions $\bm{x}_{a,{\rm cl}}=\bmX_{\rm cl} + \bar{\bmx}_{a,{\rm cl}}$ from the relative positions $\bar{\bmx}_{a,{\rm cl}}$ to the COM. Hence, we correctly reproduce the gravitational waveform emitted from the generic scattering trajectories of the worldlines under the two-body potential in linearised GWs. Higher-order effects of gravity can be added by extending the radiation potential to higher orders in $G$.

Before concluding, note that our analysis has been based on the initial condition for scattering orbits. However, as we mentioned in footnote~\ref{footnote}, the difference between scattering and bound state wavefunctions is the boundary condition of $\sigma_{\bmp}$, so the same analysis can be applied to the bound state wavefunctions. Since our last formula \eqref{T_formula} only requires the classical trajectory of the particles, we can calculate the wave emission from a bound orbit by the same formula \eqref{T_formula} by simply changing the classical trajectory from a scattering orbit to the bound orbit.


\section{Conclusion and outlook}
\label{sec:conclusion}
In this paper, we have developed a formalism to calculate non-perturbative inelastic scattering amplitudes under the classical limit and applied it to the waveform calculation emitted from conservative two-body dynamics. The idea and main results are summarised in Sec.~\ref{sec:intro}. To summarise, we have found that the generation of GWs can be computed by integrating the radiation potential along the classical trajectory governed by the post-Minkowskian two-body potential. The radiation potential can be extracted from perturbative scattering amplitudes via Born subtraction. This may be regarded as an extension of the EFT matching of the gravitational potential to radiative processes, formulated through Feshbach’s projector formalism and resummed amplitudes.
There are various directions to develop further, which we list as follows.
\begin{itemize}
    \item {\bf Higher-order radiation potential.}
    We have calculated the radiation potential at leading order and confirmed that our formula is equivalent to the GW emission from the energy-momentum tensor of two worldlines. It is natural to extend this calculation to higher orders in $G$, where general-relativistic corrections, e.g., graviton self-interactions, should come in (see the top panel of Fig.~\ref{fig:resum_amp}). Along this line, it would be interesting to explore if there is a more efficient way to extract the radiation potential or the waveform than using the Born subtraction with the relativistic Hamiltonian, similarly in the case of the two-body dynamics~\cite{Damour:2017zjx, Bern:2019crd, Kalin:2019rwq, Bjerrum-Bohr:2019kec, Cristofoli:2020uzmZ}.
    \item {\bf Centre-of-mass dynamics.}
    As is well known, in general, the centre-of-mass (COM) position is no longer constant in relativistic two-body problems, and its motion also contributes to wave emission. This effect is also evident in our formalism, where COM motion arises from deviations from the COM frame due to the presence of the graviton state. This motivates us to extract information about the COM motion in addition to the relative motion. This is particularly relevant when estimating constants of motion for bound systems, which can be approached using our formalism. Once the COM coordinates are determined, one can readily consider circular orbits and extract key quantities such as their binding energy \cite{Bernard:2017ktp,Foffa:2012rn,Bini:2017wfr}, which, together with the radiated flux, contribute to the measured phase of gravitational-wave signals. 
   \item {\bf Gravitational wave propagation.} 
    In this paper, we have only implicitly dealt with the graviton wavefunction $\Psi^-_{\bmk}$ and not discussed the precise justification of the factorisation \eqref{three-body_state} nor the equation of motion to determine $\Psi^-_{\bmk}$. It is desirable to sort out these details; for example, see~\cite{Correia:2024jgr, Caron-Huot:2025tlq} for the recent use of the Born series to wave dynamics. Meanwhile, since our general formula is quite close to the one obtained from the black hole perturbation~\cite{Aoki:2024boe}, one might be able to approximate it by the solution to the Teukolsky equation, especially if one is interested in black hole mergers commented below. It would also be interesting to use our formalism to compare predictions that Sachs's peeling property is violated at leading order in the post-Minkowskian expansion, as shown recently in \cite{DeAngelis:2025vlf}.

    \item {\bf Non-conservative dynamics.} 
    The effective potentials of Feshbach are naturally complex and can describe inelastic (non-conservative) effects during scattering processes. In fact, radiation reaction~\cite{DiVecchia:2020ymx, Damour:2020tta,DiVecchia:2021ndb,DiVecchia:2021bdo,Elkhidir:2023dco}, black hole absorption~\cite{Aoude:2023fdm, Jones:2023ugm, Chen:2023qzo, Aoude:2024jxd, Bautista:2024emt, Gatica:2025uhx}, and even the Hawking radiation~\cite{Aoki:2025ihc} can be understood as the existence of additional inelastic channels.\footnote{See also~\cite{Aoude:2024sve, Aoude:2025jvt, Ilderton:2025aql, Carrasco:2025bgu} for the recent developments of other aspects of the amplitudes and Hawking radiation.} This is another advantage to adopt the notion of the effective potential in the Schr\"{o}dinger equation, instead of using the notion of the classical potential. As we mentioned at the end of Sec.~\ref{sec:KMOCwaveform}, nevertheless, we should modify the derivation of the waveform in the presence of additional channels. It would be intriguing to study whether we can correctly describe the wave emission from non-conservative dynamics by extending the method of this paper.
    \item {\bf Black hole merger.} Finally, one of the important observables in the current GW experiments is the waveforms from coalescences of two black holes. In the amplitude perspective, the merger is understood as a complete absorption process~\cite{Aoki:2024boe}, which can be realised by adding an imaginary part to the effective potential at a finite $r$. Also, for the black hole mergers, strong gravity effects should be crucial, requiring a resummation of the effective potentials, i.e., resumming the series expansions in the top panel of Fig.~\ref{fig:resum_amp}. Since our general formula is valid for all orders in $G$, it may be possible to explore intersections with other strong-gravity methods, such as black hole perturbation/self-force expansion~\cite{Pound:2021qin,Bern:2025zno,Barack:2023oqp,Cheung:2024byb,Kosmopoulos:2023bwc,Brammer:2025rqo} and the effective one-body approach~\cite{Buonanno:1998gg, Buonanno:2000ef, Damour:2012mv}.
\end{itemize}

\acknowledgments
We are grateful to Dogan Akpinar and Hyun Jeong for multiple discussions during this project and Yu-tin Huang for drawing our attention to the Sommerfeld enhancement, which led to the initiation of this project. The diagrams in this paper were drawn with the help of \texttt{TikZ-FeynHand}\, \cite{Ellis:2016jkw}.  The work of K.A. was supported by JSPS KAKENHI Grant Nos.~JP24K17046 and JP24KF0153. A.C. was supported by JSPS KAKENHI Grant No. JP24KF0153.


\appendix
\section{Relativistic WKB approximation}
\label{sec:saddle}
This Appendix derives \eqref{eq_bmp}-\eqref{WKB_wf2} from \eqref{Sch_hbar}. A similar analysis can be found in \cite{Cea:1982rg, Cea:1983wt} for a relativistic Hamiltonian $H_{12}(\bmp,\bmx)=\sum_a\sqrt{m_a^2+\bmp^2}+V(r)$ in spherical coordinates. On the other hand, our calculations are performed in Cartesian coordinates rather than spherical coordinates, and applicable to a generic momentum-dependent Hamiltonian $H_{12}(\bmp,\bmx)$ with the dependence of the total momentum being implicit.

\subsection{Saddle-point integrals}
Let us start with calculating the integral 
\begin{align}
    I(\hbar)=\int \dd^n z F(z) e^{-\frac{i}{\hbar}f(z)}
    \,, \label{saddle_int}
\end{align}
under the saddle-point approximation in higher orders in $\hbar$ (see e.g.,~\cite{schafer1967higher, nicola2022saddle, bleistein1975asymptotic}). We use the notation like
\begin{align}
    f_I(z)=\frac{\partial f(z)}{\partial z^I}\,, \qquad  f_{IJ}(z)=\frac{\partial^2 f(z)}{\partial z^I \partial z^J}\,,
\end{align}
and denote the saddle by $z_0$:
\begin{align}
    f_I(z_0)=0
    \,.
\end{align}
We expand the functions around the saddle $z_0$,
\begin{align}
    F(z)&=F(z_0) + F_I(z_0)w^I + \frac{1}{2} F_{IJ}(z_0)w^I w^J + \cdots\,, \\
    f(z)&=f(z_0) + f_I(z_0)w^I + \frac{1}{2} f_{IJ}(z_0)w^I w^J + \frac{1}{6}f_{IJK}(z_0)w^I w^J w^K 
    \nn
    &+ \frac{1}{24}f_{IJKL}(z_0)w^I w^J w^K w^L +\cdots\,,
\end{align}
where $w:=z-z_0$. The integral \eqref{saddle_int} can then be written as
\begin{align}
    I=F(z_0)e^{-\frac{i}{\hbar}f(z_0)} \int \dd^n w\, g(w)\,e^{-\frac{i}{2\hbar} w^I f_{IJ}(z_0) w^J} 
\end{align}
which we can compute by using the Gaussian integrals. 

Let $G^{IJ}$ be the inverse of the Hessian, $G^{IJ}=f_{IJ}^{-1}$. Using the shorthand
\begin{align}
 \langle g(w)\rangle= \left(\frac{i}{2\pi \hbar}\right)^{n/2}\sqrt{\det f_{ij}} \int \dd^n w\, g(w)\,e^{-\frac{i}{2\hbar} w^I f_{IJ}(z_0) w^J} 
 \,,
\end{align}
the Gaussian integrals are given by
\begin{align}
     \langle 1 \rangle &=1 \,, \\
     \langle w^I w^J \rangle &= \frac{\hbar}{i} G^{IJ}\,, \\
     \langle w^I w^J w^K w^L \rangle &= \langle w^I w^J \rangle \langle w^K w^L \rangle + \langle w^I w^K \rangle \langle w^J w^L \rangle + \langle w^I w^L \rangle \langle w^J w^K \rangle
     \nn
     &=\left(\frac{\hbar}{i}\right)^2( G^{IJ}G^{KL}+G^{IK}G^{JL}+G^{IL}G^{JK})
     \,.
\end{align}
The general case is obtained by the Wick theorem
\begin{align}
    \langle w^{I_1} w^{I_2}\cdots w^{I_{2N+1}} \rangle &= 0
    \,, \\
    \langle w^{I_1} w^{I_2}\cdots w^{I_{2N}} \rangle &= \left(\frac{\hbar}{i} \right)^N \sum_{\text{all pairing}} G^{I_{\alpha_1} J_{\beta_1}} \cdots G^{I_{\alpha_n} J_{\beta_n}}\,.
\end{align}

We compute \eqref{saddle_int} up to the sub-leading order in the $\hbar$ expansion. This requires to evaluate the function $g(w)$ up to the following order
\begin{align}
    g&=1 + \frac{1}{2F}F_{IJ}w^I w^J -\frac{i}{\hbar}\left( \frac{1}{6F}F_I f_{JKL} +\frac{1}{24}f_{IJKL}\right) w^I w^J w^K w^L
    \nn
    &  + \left(\frac{i}{\hbar}\right)^2 \frac{1}{72}f_{IJK}f_{LMN} w^I w^J w^K w^L w^M w^N +\cdots
\end{align}
where $\cdots$ are terms that vanish or become higher order in $\hbar$ when performing the Gaussian integrals. Then, we obtain
\begin{align}
    I&=\left(\frac{2\pi \hbar}{i}\right)^{n/2}\frac{Fe^{-\frac{i}{\hbar}f}}{\sqrt{\det f_{IJ}}}
    \nn
    &\times \Biggl[ 1 -i\hbar \Biggl(\frac{1}{2F}F_{IJ}G^{IJ} -\frac{1}{2F}F_I f_{JKL}G^{IJ}G^{KL} - \frac{1}{8}f_{IJKL}G^{IJ}G^{KL} 
    \nn
    & \qquad \qquad\qquad+ \frac{1}{12}f_{IJK}f_{LMN}G^{IL}G^{JM}G^{KN} +\frac{1}{8}f_{IJK}f_{LMN}G^{IJ}G^{LM}G^{KN}\Biggl)
     +\mathcal{O}(\hbar^2)\Biggl]
     .
     \label{saddle_general}
\end{align}

\subsection{Relativistic Schr\"{o}dinger equation}
We apply the general result \eqref{saddle_general} to the integral form of the Schr\"{o}dinger equation; that is, $z=(\bmx',\bmp')$ and
\begin{align}
    F=\frac{1}{(2\pi \hbar)^3}\left[1 + \frac{i\hbar}{2}\partial_{\bmp'}\cdot \partial_{\bmx'} +\cdots \right]H_{12}(\bmp',\bmx')\,, \qquad f=\bmp'\cdot(\bmx'-\bmx)-\sigma_{\bmp}(\bmx')
    \,.
\end{align}
The Hessian and its inverse are given by
\begin{align}
    f_{IJ}=
    \begin{pmatrix}
    -\partial^2_{\bmx'} \sigma_{\bmp} & \bm{1} \\
    \bm{1} &  \bm{0}\\
    \end{pmatrix}
    \,, \qquad
    G^{IJ}=
    \begin{pmatrix}
       \bm{0} & \bm{1} \\
        \bm{1} &  \partial^2_{\bmx'} \sigma_{\bmp}  \\
    \end{pmatrix}
    \,.
\end{align}
Therefore, the saddle-point approximation of \eqref{Sch_hbar} gives
\begin{align}
    E_{12}-H_{12}+i\hbar \left(\frac{1}{2}\frac{\partial^2 H_{12}}{\partial p'_i \partial p'_j} \frac{\partial^2 \sigma_{\bmp}}{\partial x^i \partial x^j} +  \frac{1}{2}\frac{\partial^2 H_{12}}{\partial p'_i \partial x^i} + \frac{1}{A_{\bmp}}\frac{\partial H_{12}}{\partial p'_i} \frac{\partial A_{\bmp}}{\partial x^i} \right) +\mathcal{O}(\hbar^2)=0
    \,,
    \label{Sch_saddle}
\end{align}
under $\bmp'=\partial_{\bmx} \sigma_{\bmp}(\bmx)=\bmpb(\bmx)$. In the following, we keep using $\bmp'$ (and set $\hbar=1$) for notational simplicity, but it is understood that it is evaluated at $\bmp'=\bmpb(\bmx)$. 

We define the phase $\sigma_{\bmp}$ as the solution to the leading part of \eqref{Sch_saddle} in order to split it into a set of equations
\begin{align}
E_{12}(\bmp)-H_{12}(\bmp',\bm{x})&=0
\,, \label{WKB_Sch1} \\
\frac{1}{2}\frac{\partial^2 H_{12}}{\partial p'_i \partial p'_j} \frac{\partial p'_i}{\partial x^j} + \frac{1}{2} \frac{\partial^2 H_{12}}{\partial p'_i \partial x^i} + \frac{1}{A_{\bmp}}\frac{\partial H_{12}}{\partial p'_i} \frac{\partial A_{\bmp}}{\partial x^i}&=0
\,, \label{WKB_Sch2}
\end{align}
where $p'_i$ satisfies
\begin{align}
    \frac{\partial p'_i}{\partial x^j}=\frac{\partial p'_j}{\partial x^i}=\frac{\partial^2 \sigma_{\bmp}}{\partial x^i \partial x^j}
    \,.
\end{align}
Eq.~\eqref{WKB_Sch2} can be written in the form of the continuity equation \eqref{eq_continuity} by introducing the bulk velocity $\bm{v}_{\rm bulk}:=\left. \frac{\partial H_{12}}{\partial \bmp}\right|_{\bmp=\bmp_{\rm bulk}(\bmx)}$. Hence, Eqs.~\eqref{WKB_Sch1} and \eqref{WKB_Sch2} are the well-known set of equations under the WKB approximation, but are valid for a generic relativistic Hamiltonian $H_{12}=H_{12}(\bmp,\bmx)$.

To solve \eqref{WKB_Sch2}, we use the following identities obtained by differentiating \eqref{WKB_Sch1}:
\begin{align}
    \frac{\dd}{\dd p_{i}} \eqref{WKB_Sch1} =0~&\implies~\frac{\partial E_{12}}{\partial p_{i}} - \frac{\partial H_{12}}{\partial p'_j} \frac{\partial p'_j}{\partial p_{i}} =0
    \,, \\
    \frac{\dd^2}{\dd p_{i} \dd x^j} \eqref{WKB_Sch1} =0~&\implies~\left( \frac{\partial^2 H_{12}}{\partial p'_k \partial p'_l} \frac{\partial p'_l}{\partial x^j}
    + \frac{\partial^2 H_{12}}{\partial p'_k \partial x^j}\right)\frac{\partial p'_k}{\partial p_{i}} 
    + \frac{\partial H_{12}}{\partial p'_k} \frac{\partial^2 p'_k}{\partial p_{i} \partial x^j} =0
    \,.
\end{align}
Multiplying the inverse $\left( \frac{\partial p'_j}{\partial p_{i}} \right)^{-1}$, the second equation yields
\begin{align}
    0&=\frac{\partial^2 H_{12}}{\partial p'_k \partial p'_l} \frac{\partial p'_l}{\partial x^j}
    + \frac{\partial^2 H_{12}}{\partial p'_k \partial x^j} + \left( \frac{\partial p'_k}{\partial p_{i}} \right)^{-1}\frac{\partial H_{12}}{\partial p'_l} \frac{\partial^2 p'_l}{\partial p_{i} \partial x^j} 
    \nn
    &=\frac{\partial^2 H_{12}}{\partial p'_k \partial p'_l} \frac{\partial p'_l}{\partial x^j}
    + \frac{\partial^2 H_{12}}{\partial p'_k \partial x^j} + \left( \frac{\partial p'_k}{\partial p_{i}} \right)^{-1}\frac{\partial H_{12}}{\partial p'_l} \frac{\partial^2 p'_j}{\partial p_{i} \partial x^l} 
    \,.
\end{align}
Then, its trace can be written by
\begin{align}
    \frac{1}{2}\left( \frac{\partial^2 H_{12}}{\partial p'_i \partial p'_j} \frac{\partial p'_i}{\partial x^j}
    + \frac{\partial^2 H_{12}}{\partial p'_i \partial x^i} \right)
    + \frac{1}{\sqrt{\det \partial_{\bmp} \partial_{\bm{x}} \sigma_{\bmp}(\bm{x})}} \frac{\partial H_{12}}{\partial p'_i} \frac{\partial}{\partial x^i} \sqrt{ \det \partial_{\bmp} \partial_{\bm{x}} \sigma_{\bmp}(\bm{x})}=0
    \,,
\end{align}
which exactly agrees with \eqref{WKB_Sch2} under $A_{\bmp}\propto \sqrt{\det \partial_{\bmp} \partial_{\bm{x}} \sigma_{\bmp}(\bm{x})}$. The proportional constant is fixed to be unity by imposing the boundary condition $A_{\bmp} \to 1$ as $\sigma_{\bmp}\to \bmp \cdot \bmx$.

Note that the analysis so far does not require the Hamiltonian to be Hermitian. Let us now assume a Hermitian Hamiltonian (conservative system) and confirm that the wavefunction satisfies completeness 
\begin{align}
    \hat{1}\eqcons \int \hd^3 \bmp \ket{\Psi^+_{\bmp}}\bra{\Psi^+_{\bmp}} \quad \Rightarrow \quad
    \delta^{(3)}(\bmx'-\bmx)\eqcons \int \hd^3\bmp [\Psi^+_{\bmp}(\bmx')]^* \Psi^+_{\bmp}(\bmx)
    \label{WKB_completness}
\end{align}
and orthogonality
\begin{align}
    \hdelta^{(3)}(\bmp'-\bmp)\eqcons \braket{\Psi^+_{\bmp'}|\Psi^+_{\bmp}}\quad \Rightarrow \quad
    \hdelta^{(3)}(\bmp'-\bmp)\eqcons \int \dd^3\bmx [\Psi^+_{\bmp'}(\bmx)]^* \Psi^+_{\bmp}(\bmx)
    \,,
\end{align}
within the WKB approximation. We only compute the former, as the latter can be verified similarly. Due to the rapid oscillation of the wavefunction, the integral $\int \hd^3\bmp [\Psi^+_{\bmp}(\bmx')]^* \Psi^+_{\bmp}(\bmx)$ must have support only around $\bmx \simeq \bmx'$. Assuming $\sigma_{\bmp}$ is real, we obtain
\begin{align}
    \int \hd^3\bmp [\Psi^+_{\bmp}(\bmx')]^* \Psi^+_{\bmp}(\bmx) &\simeq \int \hd^3 \bmp |A_{\bmp}(\bmx)|^2 e^{-i \partial_{\bmx}\sigma_{\bmp}(\bmx)\cdot (\bmx'-\bmx)}
    \nn
    &= \int \hd^3 \bmp_{\rm bulk} \frac{|A_{\bmp}(\bmx)|^2}{|\det \partial_{\bmp} \partial_{\bmx}\sigma_{\bmp}(\bmx)|} e^{-i\bmp_{\rm bulk}\cdot (\bmx'-\bmx)}
    \,.
\end{align}
Hence, one sees that $A_{\bmp}=\sqrt{\det \partial_{\bmp} \partial_{\bmx}\sigma_{\bmp}(\bmx)}$ yields the correct normalisation of the wavefunction. Conversely, by imposing \eqref{WKB_completness}, one can obtain $A_{\bmp}=\sqrt{\det \partial_{\bmp} \partial_{\bmx}\sigma_{\bmp}(\bmx)}$ without going through evaluating the saddle-point integral up to the sub-leading order, although this cannot be applied to the non-Hermitian case.

Finally, we briefly comment on the role of the derivative terms in the integral form of the Schr\"{o}dinger equation \eqref{Sch_position}. They can be understood as arising from the non-commutativity of momenta and positions, which becomes manifest by moving the integral form to the differential form. Note that this is different from canonical quantisation, where there is a well-known ambiguity of the ordering of $\bmp$ and $\bmx$ when they are promoted to operators. Here, there is no such ambiguity because we just move a representation of the Schr\"{o}dinger equation to another representation. Let us consider a simple example $V_{\rm PM}=c_1\bmx\cdot \bmp$. In such a case, we obtain
\begin{align}
   &\int \hd^3 \bmp' \dd^3 \bmx' e^{-i\bmp'\cdot (\bmx'-\bmx)} \Psi^{+}_{\bm{p}}(\bmx') \left[1 + \frac{i}{2}\partial_{\bmp'}\cdot \partial_{\bmx'} +\cdots \right] V_{\rm PM}(\bmp',\bmx') 
   \nn
   &=c_1 \int \hd^3 \bmp' \dd^3 \bmx' e^{-i\bmp'\cdot (\bmx'-\bmx)}  \left[ \hat{\bmp}'\cdot \bmx'  + \frac{i}{2} \right] \Psi^{+}_{\bm{p}}(\bmx') 
   \nn
   &=\frac{c_1}{2}\left[\bmx\cdot \hat{\bmp}+\hat{\bmp}\cdot \bmx \right] \Psi^{+}_{\bm{p}}(\bmx)
\end{align}
with $\hat{\bmp}'=-i\partial_{\bmx'}$ and $\hat{\bmp}=-i\partial_{\bmx}$. We correctly get the self-adjoint potential $\frac{c_1}{2}(\bmx\cdot \hat{\bmp}+\hat{\bmp}\cdot \bmx)$ thanks to the derivative term on the first line.

\section{All order impulse for conservative dynamics}
\label{sec:impulse}
In this section, we compute the impulse
\begin{align}
    \Delta p_1^{\mu}:=\braket{\Psi|\hat{S}^{\dagger} \hat{P}_1^{\mu}\hat{S}|\Psi}-\braket{\Psi| \hat{P}_1^{\mu}|\Psi}
\end{align}
and confirm the agreement with classical physics for a generic conservative two-body potential. Recall that conservative dynamics implies (i) the two-body potential $V_{\rm PM}(\bmp,\bmx)$ and the WKB phase $\sigma_{\bmp}(\bmx)$ are real and (ii) the completeness relation can be truncated up to the two-particle state \eqref{completness_cons}. For simplicity, we work in the COM frame and focus on the spatial component. Using the truncated completeness relation and taking the classical limit, we obtain
\begin{align}
    \Delta \bmp &= \bmp_f-\bmp_i
    \,,\\
    \bmp_f&\eqcons \int\left[ \prod_{a=1,2} \hd^4 q_{a} \hdelta(2p_{a} \cdot q_{a})e^{ib_a \cdot q_{a}} \right]\nn
    &\times\int \dd \Phi(p_1',p_2')\bmp' \braket{p_{1}-q_1;p_{2}-q_2|\hat{S}^{\dagger}|p_{1}';p_{2}'} \braket{p_1';p_2'|\hat{S}|p_1;p_2}
    \,, \label{final_momentum} \\
    \bmp_i&=\bmp
    \,.
\end{align}
Note that we ignore the $q^2$ term in the on-shell delta functions of \eqref{final_momentum}. This should not be confused with the statement in~\cite{Cristofoli:2021jas} that the $q^2$ term is necessary to compute the impulse beyond the leading order. In our case, $q$ is the {\it momentum mismatch} of the in momenta of the wavepackets, while $q$ in \cite{Cristofoli:2021jas} is the {\it momentum transfer} between the in and out momenta. We assume $q_a=\mathcal{O}(\hbar)$, but do not assume $p_a-p_a' = \mathcal{O}(\hbar)$ consistently with~\cite{Cristofoli:2021jas}.

Using the interacting states and the variables introduced in \eqref{PQ_var1} and \eqref{PQ_var2}, the final momentum is 
\begin{align}
    \bmp_f &= \int\left[ \prod_{a=1,2} \hd^4 q_{a} \hdelta(2p_{a} \cdot q_{a})e^{ib_a \cdot q_{a}} \right] 4E_1 E_2\,\hdelta^{(3)}(\bm{Q})
    \int \hd^3 \bmp' \,\bmp' \braket{\Psi^+_{\bmp-\bmq}|\Psi^-_{\bmp'}}\braket{\Psi^-_{\bmp'}|\Psi^+_{\bmp}}
     \nn
    &= \int \hd^3 \bmq \hd^3 \bmp' e^{-i\bm{b}\cdot \bmq} \bmp' \braket{\Psi^+_{\bmp-\bmq}|\Psi^-_{\bmp'}}\braket{\Psi^-_{\bmp'}|\Psi^+_{\bmp}}
    \,.
\end{align}
We have performed $q^0, Q^0,$ and $\bm{Q}$ integrals with the help of the two on-shell delta functions $\hdelta(2p_{a} \cdot q_{a})$ and the delta function for spatial conservation $\hdelta^{(3)}(\bm{Q})=\hdelta^{(3)}(\bmq_1+\bmq_2)$. Note that there is only the spatial delta function on the first line because the energy conservation is hidden in the interacting states (see \eqref{S-matrix}). It can be compared with the waveform \eqref{W_delta}, where one on-shell delta function remains, generating the time integral. Their difference can be understood by the distinct nature of impulse and waveform: the impulse is, by definition, only sensitive to the initial and final momenta, whereas the waveform is sensitive to the detailed bulk dynamics of massive bodies. For the former, we don't need time if the scattering data is given, which, in our case, is encoded in the wavefunctions. 

Let's make it more concrete. The WKB wavefunction gives
\begin{align}
    \bmp_f &=\int \hd ^3\bmq \hd^3 \bmp' \dd^3 \bmx \dd^3 \bmx' e^{-i\bm{b}\cdot \bmq} \bmp' 
    \nn
    &\qquad \qquad \times A^*_{\bmp-\bmq}(\bmx') A_{\bmp'}(\bmx') A^*_{\bmp'}(\bmx) A_{\bmp}(\bmx) e^{i(\sigma^+_{\bmp}(\bmx)+\sigma^-_{\bmp'}(\bmx')-\sigma^+_{\bmp-\bmq}(\bmx')-\sigma^-_{\bmp'}(\bmx))}
    \nn
    &\simeq \int \hd ^3\bmq \hd^3 \bmp' \dd^3 \bmx \dd^3 \bmx' e^{-i\bmq \cdot (\bm{b}- \partial_{\bmp}\sigma^{+}_{\bmp}(\bmx')) } \bmp' 
    \nn
    &\qquad \qquad \times A^*_{\bmp}(\bmx') A_{\bmp'}(\bmx') A^*_{\bmp'}(\bmx) A_{\bmp}(\bmx) e^{i(\sigma^+_{\bmp}(\bmx)+\sigma^-_{\bmp'}(\bmx')-\sigma^+_{\bmp}(\bmx')-\sigma^-_{\bmp'}(\bmx))}
    \nn 
    &=\int \hd^3 \bmp' \dd^3 \bmx \dd^3 \bmx' \delta^{(3)}(\bm{b}- \partial_{\bmp}\sigma^{+}_{\bmp}(\bmx'))\, \bmp'
    \nn
    &\qquad \qquad \times A^*_{\bmp}(\bmx') A_{\bmp'}(\bmx') A^*_{\bmp'}(\bmx) A_{\bmp}(\bmx) e^{i(\sigma^+_{\bmp}(\bmx)+\sigma^-_{\bmp'}(\bmx')-\sigma^+_{\bmp}(\bmx')-\sigma^-_{\bmp'}(\bmx))}
\end{align}
for small $\bmq$. Here, the superscripts of $\sigma^{\pm}_{\bmp}$ denote the in/out boundary conditions. The delta function localises $\bmx'$ to be $\bmx'=\bm{b}_{\rm eik}$ with $\bm{b}_{\rm eik}$ denoting the solution to the delta function. The saddle of $\bmx$ and $\bmp'$ integrals is 
\begin{align}
    \partial_{\bmx}\sigma^+_{\bmp}(\bmx)=\partial_{\bmx}\sigma^-_{\bmp'}(\bmx)\,, \qquad \partial_{\bmp'}\sigma^-_{\bmp'}(\bmx)=\partial_{\bmp'}\sigma^-_{\bmp'}(\bmx')
    \,,
\end{align}
the latter of which can be solved by $\bmx=\bmx'$. Hence, the saddle-point approximation, together with the support of the delta function, yields
\begin{align}
    \bmp_f=\bmp'
\end{align}
with $\bmp'$ being a solution to
\begin{align}
    \bm{b} - \partial_{\bmp}\sigma^+_{\bmp}(\bm{b}_{\rm eik})=0\,, \qquad
    [\partial_{\bmx}\sigma^+_{\bmp}(\bmx)-\partial_{\bmx}\sigma^-_{\bmp'}(\bmx)]_{\bmx = \bm{b}_{\rm eik}} =0
    \,.
    \label{impulse}
\end{align}
To understand that this reproduces the correct classical physics, we identify $\sigma^{\pm}_{\bmp}$ with Hamilton's characteristic functions of classical physics. The trajectory equations are
\begin{align}
    \bm{b}+\partial_{\bmp}E_{12} t -\partial_{\bmp} \sigma_{\bmp}^{\pm}(\bmx)=0
    \,.
\end{align}
According to the boundary conditions $\sigma^{\pm}_{\bmp}$, the solution for $\pm$ corresponds to the classical trajectory with the momentum $\bmp$ at $t=\mp \infty$ and with the impact parameter $\bm{b}$. The first condition in \eqref{impulse} is thus understood as determining the position of the two particles at $t=0$, and the second one provides the matching condition of the in/out trajectories at $t=0$; the in/out trajectories are at the same position $\bmx=\bm{b}_{\rm eik}$ and have the same momentum $\partial_{\bmx}\sigma^+_{\bmp}=\bmp_{\rm bulk}(\bm{b}_{\rm eik})=\partial_{\bmx}\sigma^-_{\bmp'}$ at $t=0$. As a result, the solution $\bmp'$ to \eqref{impulse} agrees with the final momentum of the two-body dynamics for the initial momentum $\bmp$ and the impact parameter $\bm{b}$.
\newpage

\bibliographystyle{JHEP}
\bibliography{biblio}

\end{document}